%% file: Main.tex
\PassOptionsToPackage{table}{xcolor}
\documentclass[acmsmall]{acmart}

\AtBeginDocument{%
  }

\usepackage{tikz}
\usepackage{xspace}
\usepackage{enumitem}
\usepackage{multirow}
\usepackage{array}
\usepackage{threeparttable}
\usepackage{booktabs}
\usepackage{tabularx}
\usepackage{pifont}
\usepackage{url}
\usepackage{float}
\usepackage{graphicx}
\usepackage{wrapfig2}
\usepackage{floatflt}
\usepackage[most, breakable]{tcolorbox}
\usepackage{subcaption}

\input{macro}

\newcommand{\bench}{SIExBench\xspace}

\newcommand{\infoex}{SIEx\xspace}

\setcopyright{acmlicensed}
\copyrightyear{2018}
\acmYear{2018}
\acmDOI{XXXXXXX.XXXXXXX}
\acmISBN{978-1-4503-XXXX-X/2018/06}

\begin{document}

\title[\bench]{{\bench}: A Benchmark for Automated Repair of Sensitive Information Exposure}

\author{Mingming Zhang}
\email{mmzhang@buaa.edu.cn}
\affiliation{%
  \institution{Beihang University}
  \city{Beijing}
  \country{China}
}

\author{Xu Wang}
\email{xuwang@buaa.edu.cn}
\affiliation{%
  \institution{Beihang University}
  \city{Beijing}
  \country{China}
}

\author{Jian Zhang}
\authornote{Corresponding author.}
\email{zhangj\_cs@buaa.edu.cn}
\affiliation{%
  \institution{Beihang University}
  \city{Beijing}
  \country{China}
}

\author{Dihao Fan}
\email{fandihao@buaa.edu.cn}
\affiliation{%
  \institution{Beihang University}
  \city{Beijing}
  \country{China}
}

\author{Xiangxin Meng}
\email{mengxx@buaa.edu.cn}
\affiliation{%
  \institution{Beihang University}
  \city{Beijing}
  \country{China}
}

\author{Chunming Hu}
\email{hucm@buaa.edu.cn}
\affiliation{%
  \institution{Beihang University}
  \city{Beijing}
  \country{China}
}

\renewcommand{\shortauthors}{Zhang et al.}

\begin{abstract}
Modern software systems routinely process credentials, personal records, and other sensitive information. A sensitive information exposure (\infoex) defect makes such information accessible to an unauthorized actor and can cause serious and often irreversible disclosure. However, prior research has focused largely on detecting such exposures rather than repairing the underlying defects, and existing repository-level issue-resolution benchmarks do not specifically target this defect class. We present {\bench}, to our knowledge the first benchmark dedicated to automated repair of \infoex defects. {\bench} contains 161 execution-verified instances from 64 widely used Python and Java repositories, each with an issue report, buggy repository snapshot, developer patch, executable exposure and regression tests, and defect characterization.

We evaluate six repair systems with three LLMs, yielding 18 configurations, and systematically analyze their overall effectiveness, performance across defect characteristics, and failure mechanisms. The strongest configuration resolves only 39.1\% of the instances, while 40.4\% remain unresolved by all 18 configurations. More importantly, 62.1\% of unsuccessful attempts produce a valid patch that still leaves sensitive information exposed, making incomplete exposure removal a central challenge. Systems often generate plausible patches and modify relevant code without fully eliminating the exposure, indicating that effective repair must ensure that all relevant paths to sensitive-information exposure are closed. Overall, {\bench} provides a foundation for systematically evaluating and advancing automated repair of sensitive information exposure defects.
\end{abstract}



\received{20 February 2007}
\received[revised]{12 March 2009}
\received[accepted]{5 June 2009}

\maketitle

\input{1_Introduction}

\input{2_BackgroundAndRelatedWork}

\input{3_BenchmarkConstruction}

\input{4_BenchmarkStatistics}

\input{5_ExperimentalSetup}

\input{6_Evaluation}

\input{7_ThreatsToValidity}

\input{8_Conclusion}

\section*{Data Availability}
We will publicly release {\bench}, along with the evaluation scripts and results, to support reproducibility and future research on \infoex repair.

\bibliographystyle{ACM-Reference-Format}
\bibliography{reference}

\end{document}

%% file: macro.tex
\newlength{\databarw}
\newcommand{\databar}[4]{%
  \makebox[\databarw][l]{%
    \makebox[0pt][l]{\color{#3}\rule[-0.5ex]{\dimexpr#1\databarw/#2\relax}{1.7ex}}%
    \makebox[\databarw][r]{#4}}}

\newtcolorbox{rqsummary}{
  enhanced, breakable,
  colback=black!7, colframe=black!75,
  frame hidden, boxrule=0pt, arc=0pt, outer arc=0pt,
  borderline west={2.2pt}{0pt}{black!75},
  left=6pt, right=5pt, top=4pt, bottom=4pt,
  before skip=6pt, after skip=6pt,
}

\newcommand{\dingnum}[1]{%
  \ifcase#1
    \or \ding{182} %
    \or \ding{183} %
    \or \ding{184} %
    \or \ding{185} %
    \or \ding{186} %
    \or \ding{187} %
    \or \ding{188} %
    \or \ding{189} %
    \or \ding{190} %
    \or \ding{191} %
  \fi
}

%% file: 1_Introduction.tex
\section{Introduction}
\label{sec:introduction}

Modern software systems routinely process credentials, personal records, and other sensitive information. When a software defect makes such information accessible to an actor who is not explicitly authorized to access it, it creates a \emph{sensitive information exposure defect} (CWE-200)~\cite{mitreCWE200Exposure}, which we refer to as a \infoex defect. Such exposure is particularly consequential because disclosure is generally irreversible: once sensitive information is logged, returned to an unintended client, or stored in an accessible location, a later patch cannot retract what has already been observed. Exposed credentials and personal records may also enable subsequent attacks, while software vulnerabilities impose substantial economic costs~\cite{anwar2020measuring,cybersecurityventuresCybercrimeCost}. Figure~\ref{fig:infoex-examples} shows two real-world examples from our benchmark. In Redisson, a connection password is exposed through exception messages and logs, while in Home Assistant, a device location is exposed to another user's UI. Together, these examples show that the same security consequence can arise from different root causes. Redisson directly emits sensitive information, whereas Home Assistant exposes it through incorrect program logic. Repairing \infoex defects therefore requires not only identifying sensitive information, but also reasoning about the underlying cause and how the information becomes observable.

Prior work on information exposure has largely focused on detection and diagnosis rather than repair. Static and dynamic taint analyses track sensitive information through programs~\cite{arzt_flowdroid_2014,enck_taintdroid_2014}, while other techniques target specific exposure channels, including network traffic, application logs, and server-side disclosure~\cite{ren_recon_2016,chen_comprehensive_2024,koch_semi-automated_2017}. More recently, LLM-based approaches such as Argus~\cite{wang_argus_2025} and SIExVulTS~\cite{katz_siexvults_2025} directly detect CWE-200 sensitive-information leakage. In contrast, automated program repair has mainly targeted general software bugs~\cite{meng2026agent4bugfixing,yang2025APRsurvey} and broader security vulnerabilities~\cite{li2025sokAVR}, leaving \infoex defects largely unexplored. Existing issue resolution benchmarks do not fill this gap. SWE-bench and its successors~\cite{jimenez2023swebench,lee2025secBench,zan2026multi,deng2025swebenchproaiagents,guo2025omnigirl,zhong_backportbench_2025} neither specifically curate \infoex defects nor provide dedicated repair evaluation.  Moreover, generic issue-resolution benchmarks do not identify the sensitive information, exposure channel, and unauthorized actor that characterize SIEx defects, limiting systematic analysis of repair behavior and failure causes.%
The closest existing benchmark, SecLogging~\cite{yuan2026towards}, covers security issues in logging code, including sensitive-information exposure, but is limited to logging-related defects.

To address this gap, we present {\bench}, to our knowledge the first benchmark dedicated to automated repair of information exposure defects. {\bench} contains 161 execution-verified instances from 64 widely used Python and Java repositories. Each instance includes an issue report, buggy repository snapshot, developer patch, executable exposure tests that verify that sensitive information is no longer observable, and regression tests that preserve existing behavior. 
We further characterize each defect along six dimensions, enabling fine-grained analysis of repair effectiveness.

We evaluate six repair systems with three LLMs, yielding 18 configurations, and analyze effectiveness, performance across defect characteristics, and failure mechanisms. Current systems leave substantial room for improvement: the strongest configuration resolves only 39.1\% of {\bench}, while 40.4\% of instances are unresolved by all 18 configurations. More importantly, failure is dominated not by the absence of a patch, but by incomplete exposure removal: 62.1\% of unsuccessful attempts produce a valid patch that leaves sensitive information exposed. This reveals a challenge of \infoex repair. Systems often generate plausible patches and modify relevant code without fully eliminating the exposure. Effective repair therefore requires reasoning about whether all relevant paths through which sensitive information can become observable have been closed.

In summary, we make the following contributions.

\begin{itemize}

\item \textbf{Problem Formulation.} We formulate automated repair of sensitive information exposure as a distinct repair task, requiring reasoning about how sensitive information becomes observable to unauthorized actors across diverse root causes and exposure mechanisms.

\item \textbf{Benchmark.} We introduce {\bench}, to our knowledge the first benchmark dedicated to automated \infoex repair. It comprises 161 execution-verified instances from 64 Python and Java repositories and provides exposure-oriented test oracles, regression tests, and fine-grained defect characterization.

\item \textbf{Extensive Evaluation.} We evaluate six repair systems with three LLMs, yielding 18 configurations, and systematically analyze overall repair effectiveness, performance across defect characteristics, and failure mechanisms. Our results show that incomplete exposure removal dominates unsuccessful repairs and reveal that effective \infoex repair requires reasoning beyond patch generation to fully eliminate exposure paths.

\end{itemize}

%% file: 2_BackgroundAndRelatedWork.tex
\section{Background and Related Work}
\label{sec:background}

\subsection{Sensitive Information Exposure Defects}
\label{subsec:exposure-defects}

\begin{figure}[t]
  \centering
  \includegraphics[width=0.95\textwidth]{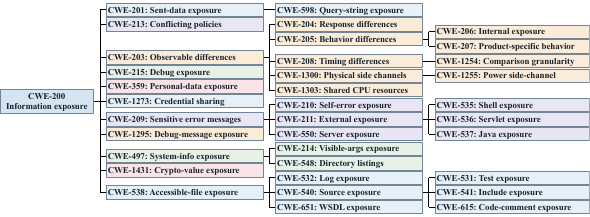}
  \Description{The CWE-200 descendant tree used to define the benchmark scope.}
  \caption{CWE-200 and its descendants; names are abbreviated.}
  \label{fig:cwe-tree}
\end{figure}

\begin{wrapfigure}{r}{0.5\linewidth}
  \centering
  \includegraphics[width=\linewidth]{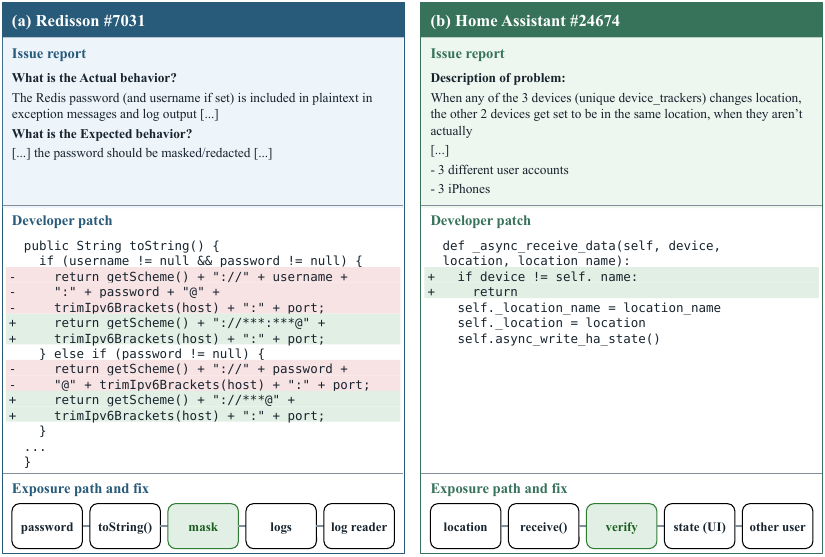}
  \Description{Two \bench instances with their issue excerpt, developer patch, and exposure path; the fix is shown in green.}
  \caption{Two \bench{} instances: issue excerpt, developer patch, and exposure path, with the fix in green.}
  \label{fig:infoex-examples}
\end{wrapfigure}
A \emph{sensitive information exposure defect} (\infoex defect) is a software defect that exposes sensitive information to an actor not explicitly authorized to access it, corresponding to CWE-200~\cite{mitreCWE200Exposure}. Figure~\ref{fig:cwe-tree} shows CWE-200 and its descendants. Our scope also includes defects whose root cause lies outside this hierarchy but whose consequence is a sensitive information exposure. Specifically, we define \infoex defects as defects whose resultant weakness falls within the CWE-200 hierarchy, without imposing any restriction on their primary weakness.\footnote{We follow the CWE Glossary definitions of \href{https://cwe.mitre.org/documents/glossary/index.html\#Primary\%20Weakness}{\emph{Primary Weakness}} and \href{https://cwe.mitre.org/documents/glossary/index.html\#Resultant\%20Weakness}{\emph{Resultant Weakness}}.}
Such defects are particularly consequential because disclosure is generally irreversible: a later patch can prevent further exposure but cannot retract information already observed. Exposed sensitive data may also enable subsequent attacks~\cite{mitreCWE200Exposure}. Given the substantial economic impact of software vulnerabilities and cybercrime~\cite{anwar2020measuring,cybersecurityventuresCybercrimeCost}, repairing these defects is an important software security problem~\cite{li2025sokAVR}.

Figure~\ref{fig:infoex-examples} shows two representative instances. In Redisson (\href{https://github.com/redisson/redisson/issues/7031}{issue \#7031}), a connection password reaches exception messages and logs through \texttt{RedisURI.toString()}, and the patch masks it before emission. In Home Assistant (\href{https://github.com/home-assistant/core/issues/24674}{issue \#24674}), a device location is incorrectly written to another device's state and exposed through the UI; the patch rejects mismatched device updates. The former is a \emph{direct exposure}, where sensitive data is emitted directly, whereas the latter is a \emph{resultant exposure}, where exposure follows from an underlying logic error.

\subsection{Sensitive Information Exposure Detection}
\label{subsec:exposure-detection}

Research on \infoex defects has focused primarily on \emph{detection} rather than repair. Static taint analyses such as FlowDroid~\cite{arzt_flowdroid_2014}, Amandroid~\cite{wei_amandroid_2018}, IccTA~\cite{li_iccta_2015}, and EdgeMiner~\cite{cao_edgeminer_2015} track sensitive data across program and component boundaries, although their accuracy depends on configuration and target~\cite{zhang_analyzing_2022}. Dynamic approaches instead observe concrete executions, including TaintDroid~\cite{enck_taintdroid_2014}, TaintART~\cite{sun_taintart_2016}, and MutaFlow~\cite{mathis_detecting_2017}. Other work targets specific exposure settings, including privacy leaks in applications~\cite{gibler_androidleaks_2012,wang_leakdoctor_2019}, network traffic~\cite{ren_recon_2016,continella_obfuscation-resilient_2017,song_privacyguard_2015,le_antmonitor_2015}, and access-control or policy enforcement~\cite{beresford_mockdroid_2011,tripp_bayesian_2014,ferrara_morphdroid_2015}. More recent studies address application logs~\cite{chen_comprehensive_2024}, server-side oversharing~\cite{koch_semi-automated_2017}, and microservice backends~\cite{liu_detecting_2025}, while LLM-based detectors such as Argus~\cite{wang_argus_2025} and SIExVulTS~\cite{katz_siexvults_2025} directly target sensitive-information leakage and CWE-200 exposures.

Overall, prior work identifies or diagnoses information exposure rather than repairing such defects. APR research has instead focused on general software bugs~\cite{meng2026agent4bugfixing,yang2025APRsurvey} and security vulnerabilities~\cite{li2025sokAVR}, leaving \infoex defects largely unexplored as a repair target. No dedicated benchmark exists for evaluating their automated repair; {\bench} fills this gap.

\subsection{LLM-Based Repository-Level Issue Resolution}
\label{subsec:issue-resolution}

Large language models (LLMs) have substantially advanced automated program repair (APR)~\cite{xia2024chatRepair,zhang2025ReinFix,ye2025adverintent,Peng2024TypeFix,fse2025mlm4dnn,yang2026morepair,zhang2023gamma,li2025hybrid,ouyang2025knowledge,akbarpour2025collaborative,bouzenia2024repairagent}. SWE-bench~\cite{jimenez2023swebench} further catalyzed research on resolving real-world repository issues from natural-language descriptions~\cite{yang2024sweAgent,gauthier2024aider,bytedance2025TRAE,liu2024marscode,wang2024openhands,xia2025agentless,ruan2025SpecRover,tao2024magis,zhang2024autocoderover,chenYeHe2025prometheus,xia2025livesweagent}.

\emph{Agent-based} systems let LLMs interact with repositories through tools. SWE-agent~\cite{yang2024sweAgent}, OpenHands~\cite{wang2024openhands}, and mini-swe-agent~\cite{githubGitHubSWEagentminisweagent} support repository inspection, editing, and test execution with varying levels of autonomy and complexity. In contrast, \emph{workflow-based} systems use predefined stages; Agentless~\cite{xia2025agentless}, for example, separates fault localization, patch generation, and patch selection. Developer-facing coding agents such as OpenCode~\cite{githubGitHubAnomalycoopencode} and Codex~\cite{githubGitHubOpenaicodex} bring similar capabilities to practical development. We evaluate representative systems from both paradigms on {\bench} to assess their effectiveness in repairing \infoex defects.

\subsection{Benchmarks for Repository-Level Issue Resolution}
\label{subsec:issue-benchmarks}

Repository-level issue resolution is supported by a growing family of benchmarks. SWE-bench~\cite{jimenez2023swebench} established the setting with 2{,}294 issues from 12 Python repositories, pairing each issue with its fixing pull request and regression tests. Later benchmarks broaden this setting to longer-horizon tasks~\cite{deng2025swebenchproaiagents,huang2026deepswemeasuringfrontiercoding}, multiple languages~\cite{zan2026multi,yang2025swesmith}, and multimodal issues~\cite{yang2024sweBenchMultimodal,guo2025omnigirl}. Other benchmarks target specific domains or defect classes, including patch backporting~\cite{zhong_backportbench_2025}, application domains~\cite{zheng_top_2025}, language- or API-specific defects~\cite{just2014defects4j,wang_defects4c_2025,mehta2026defects4rest}, and security vulnerabilities~\cite{lee2025secBench,wei_patcheval_2025,zhu_cve-bench_nodate}.

The closest related benchmark, SecLogging~\cite{yuan2026towards}, evaluates LLM-based detection and repair of 101 manually reviewed security issues in logging code, including cases of sensitive-information exposure. However, it is restricted to logging-related issues, whereas \infoex defects may arise from diverse root causes and exposure channels. Existing benchmarks therefore do not specifically curate \infoex defects or provide a dedicated evaluation setting for their repair. {\bench} fills this gap as the first benchmark dedicated to automated repair of \infoex defects.

%% file: 3_BenchmarkConstruction.tex
\section{{\bench} Construction}
\label{sec:construction}

\begin{figure*}[t]
  \centering
  \includegraphics[width=\linewidth]{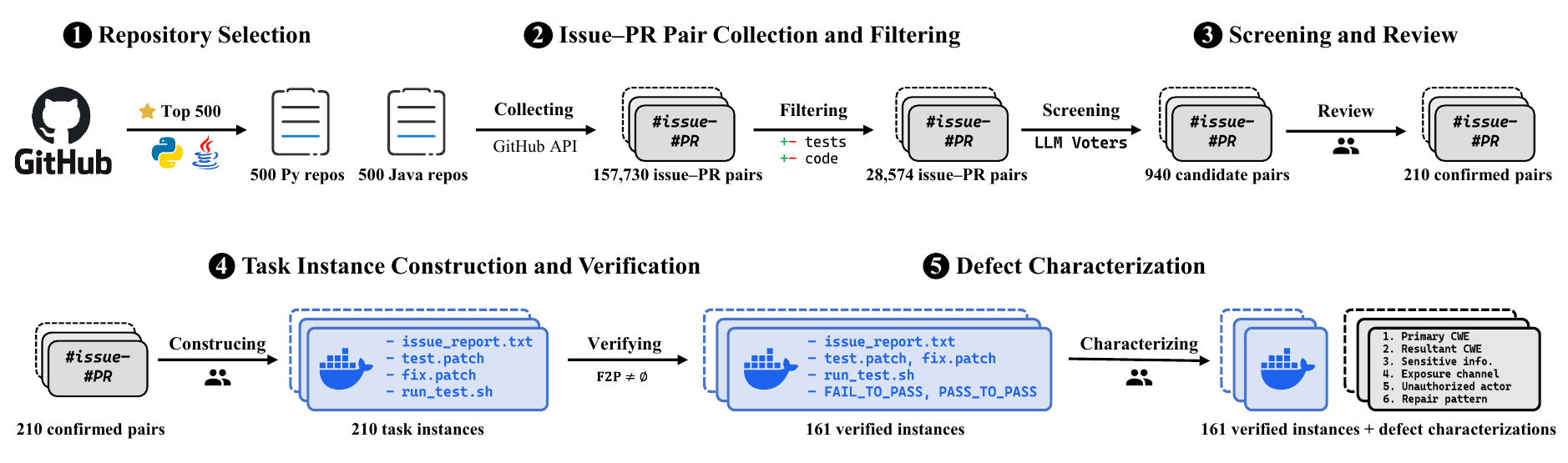}
  \caption{Overview of {\bench} construction.}
  \label{fig:pipeline}
\end{figure*}

{\bench} is constructed in six stages that progressively refine issue--PR pairs into verified benchmark instances. We \dingnum{1}select repositories, \dingnum{2}collect and filter issue--PR pairs, \dingnum{3}pre-screen and manually review candidates, \dingnum{4}construct task instances and validate them through execution, and \dingnum{5}characterize the confirmed defects along six dimensions. Figure~\ref{fig:pipeline} summarizes the pipeline, and Section~\ref{subsubsec:funnel} reports the number of candidates retained at each stage.

\subsection{Problem Definition}
\label{subsec:problem-definition}

A \infoex repair task requires generating a patch that eliminates an information exposure in a repository. The input consists of \textbf{the repository} at the commit immediately preceding the fix, referred to as the \emph{buggy commit}, and \textbf{the issue report} describing the defect; the expected output is \textbf{exactly one patch} for that snapshot. This input--output formulation follows the repository-level issue-resolution setting established by SWE-bench~\cite{jimenez2023swebench}. A task is resolved if the patch makes the exposure-revealing tests pass while preserving all previously passing tests.

\subsection{Repository Selection}
\label{subsec:repo-selection}

We sample instances from widely used open-source projects, using GitHub stars as a proxy for popularity. We focus on Python and Java, two widely used languages well represented in repository-level issue-resolution benchmarks~\cite{zan2026multi,jimenez2023swebench,yang2025swesmith}. For each language, we consider the 500 most-starred repositories and exclude private, archived, disabled, unlicensed, forked, or template repositories, as well as those that do not use GitHub Issues or pull requests. The resulting pool contains popular and legally reusable projects with established issue--PR workflows.

\subsection{Issue--PR Pair Collection and Filtering}
\label{subsec:pair-collection}

Following prior issue-resolution benchmarks~\cite{pavel-etal-2025-swe-mera,yang2025swesmith}, we require a one-to-one issue--PR mapping: each retained issue is resolved by exactly one merged PR, and each retained PR resolves exactly one issue. This establishes a clear correspondence between the issue report as the task description and the PR as its reference fix, avoiding ambiguous task boundaries when a PR addresses multiple issues or an issue is resolved across multiple PRs. We further verify each issue--PR link using recorded cross-reference events to avoid incidental mentions. Moreover, following prior APR work~\cite{xia2024chatRepair,xia2025agentless,yin2024thinkrepair}, we define the scope of our study as source-code changes and exclude fixes requiring changes to configuration, documentation, or infrastructure. Following SWE-bench~\cite{jimenez2023swebench}, we further require the fixing PR to include test changes, which provide candidate tests for constructing an executable evaluation oracle. Accordingly, we retain pairs whose changes are limited to production and test source code and modify at least one test file and one non-test source file. These criteria define our benchmark as executable, source-code-level repair and exclude issues without accompanying tests, fixes spanning multiple PRs, and fixes requiring non-code changes.

\subsection{Screening and Review}
\label{subsec:screening-review}

Rule-based filtering cannot determine whether a candidate describes a \infoex defect. We therefore use two-stage screening: recall-oriented LLM pre-screening followed by manual review. Two authors with over three years of research experience in software engineering or security independently review the retained candidates, and every instance in {\bench} is manually confirmed.

\textbf{LLM-Based Pre-screening.} We assess each issue--PR pair with DeepSeek-V4-Pro~\cite{xu2026deepseek}, providing the complete CWE-200 taxonomy and asking whether the issue describes a \infoex defect. Each pair is assessed three times and retained if any assessment identifies an exposure. This recall-oriented policy is intended to reduce false negatives, while false positives can be removed during manual review. The LLM therefore serves only as a candidate filter and makes no final inclusion decisions.

\textbf{Human Review.} Two authors independently review every retained candidate and determine whether it constitutes a \infoex defect based on three required elements: the sensitive information exposed, the exposure channel, and the unauthorized actor. A candidate is confirmed only if all three can be identified, with disagreements resolved through discussion.

\subsection{Task Instance Construction and Verification}
\label{subsec:instance-and-verification}

A confirmed defect becomes a benchmark instance only after being packaged into an executable environment and validated through execution.

\textbf{Task Instance Construction.} For each confirmed defect, we separate the merged pull request into the developer fix and test changes, and combine them with the issue report. Each instance is anchored at the buggy commit, with the issue report serving as the problem statement. We package the repository and its dependencies into a Docker image and disable network access during evaluation to prevent retrieval of external information. Figure~\ref{fig:example-instance} shows a complete instance.

\textbf{Execution-Based Verification.} We execute the tests on both the buggy and developer-fixed versions. Following SWE-bench~\cite{jimenez2023swebench}, tests that fail before and pass after the fix are classified as \texttt{FAIL\_TO\_PASS}, while those that pass in both versions are \texttt{PASS\_TO\_PASS}. During verification, we find that some developer-written \texttt{FAIL\_TO\_PASS} tests encode implementation details of the reference fix rather than the intended security outcome. For example, Figure~\ref{fig:example-instance} requires credentials to be replaced by exactly three asterisks, which would reject an alternative but secure masking strategy. We therefore manually inspect all \texttt{FAIL\_TO\_PASS} tests and rewrite such implementation-specific assertions to test the intended exposure outcome instead, e.g., that the username and password are no longer exposed through the tested behavior. For tests requiring normalization, two authors independently review each rewritten test against the issue report, developer patch, and original test to verify that it removes only implementation-specific constraints while preserving the intended exposure property. When disagreements arise, the rewritten test is revised as necessary and re-reviewed until both authors reach consensus. %
These normalized tests form the final \texttt{FAIL\_TO\_PASS} set. We retain only instances with at least one \texttt{FAIL\_TO\_PASS} test. We exclude instances whose tests (i) cannot run on the buggy commit because they depend on symbols introduced by the fix, (ii) behave identically before and after the fix, or (iii) require unavailable external services.

\textbf{Patch Evaluation Protocol.} For each candidate patch, we reset the repository to the buggy commit, apply the test changes and candidate patch, and execute the test suite. An instance is resolved only if all \texttt{FAIL\_TO\_PASS} and \texttt{PASS\_TO\_PASS} tests pass; no partial credit is awarded. As construction sanity checks, the buggy commit must fail the final \texttt{FAIL\_TO\_PASS} tests, whereas the developer fix must pass all \texttt{FAIL\_TO\_PASS} and \texttt{PASS\_TO\_PASS} tests.

\subsection{Defect Characterization}
\label{subsec:characterization}

To support fine-grained analysis, we characterize each instance along six dimensions (Table~\ref{tab:schema}). Two follow CWE weakness chains. The \emph{primary CWE} denotes the root-cause weakness, while the \emph{resultant CWE} denotes a downstream weakness caused by it.
They may differ when a \infoex defect results from another weakness; for example, a missing authorization check may be primary while the resulting exposure is resultant. When the exposure itself is the root cause, we assign the same CWE to both. Following the CWE-200 definition of information exposure, we further characterize the \emph{sensitive information} disclosed, the \emph{exposure channel} through which it becomes accessible, and the \emph{unauthorized actor} that gains access. The final dimension, \emph{repair pattern}, captures how the developer patch removes the exposure.

Two authors derive the label vocabularies through open coding. They first identify candidate labels from a sample of confirmed defects, guided by the CWE taxonomy for the two CWE dimensions and by observed exposure and repair mechanisms for the others. The labels are then refined over the remaining defects, with ambiguous categories merged and disagreements resolved through discussion. Each defect receives exactly one, most-specific applicable label per dimension. Table~\ref{tab:schema} lists the resulting categories.

\input{tables/schema}

%% file: tables/schema.tex
\definecolor{tagblue}{HTML}{2980B9}
\definecolor{taggreen}{HTML}{27AE60}
\definecolor{tagorange}{HTML}{E67E22}
\definecolor{tagpurple}{HTML}{8E44AD}

\newcommand{\tagchip}[2]{%
  \tikz[baseline=(tag.base)]{%
    \node[draw=#1, fill=#1!10, rounded corners=1.7pt, line width=0.35pt,
          inner xsep=2.8pt, inner ysep=1.0pt, outer sep=0pt,
          text height=1.55ex, text depth=0.35ex, minimum height=2.3ex,
          font=\fontsize{6.6}{7.6}\selectfont, text=black] (tag) {#2};}%
  \hspace{0.30em}\allowbreak}
\newcommand{\chips}[1]{\mbox{#1}}

\begin{table}[t]
  \centering
  \caption{The six characterization dimensions and their label vocabularies.}
  \label{tab:schema}
  \renewcommand{\arraystretch}{1.55}
  \resizebox{\linewidth}{!}{%
  \begin{tabular}{@{}l p{6.4in}@{}}
    \toprule
    \textbf{Dimension} & \textbf{Labels} \\
    \midrule
    Primary CWE & The CWE identifier of the initial weakness that constitutes the root cause of the defect and is not restricted to the CWE-200 hierarchy. \\
    Resultant CWE & The CWE identifier of the downstream exposure weakness, \textbf{which always lies within the CWE-200 hierarchy} and equals the primary CWE when the exposure itself is the root cause. \\
    Sensitive information & \chips{\tagchip{tagblue}{credential or secret}\tagchip{tagblue}{private record}\tagchip{tagblue}{business record}\tagchip{tagblue}{file content}\tagchip{tagblue}{file path}\tagchip{tagblue}{directory listing}\tagchip{tagblue}{system information}} \\
    \multirow{2}{*}{Exposure channel} & \chips{\tagchip{taggreen}{HTTP response}\tagchip{taggreen}{user interface}\tagchip{taggreen}{log}\tagchip{taggreen}{chat message}\tagchip{taggreen}{deployment config}\tagchip{taggreen}{tool result}\tagchip{taggreen}{HTTP request header}} \\
     & \chips{\tagchip{taggreen}{SQL result}\tagchip{taggreen}{error message}\tagchip{taggreen}{HTTP request}\tagchip{taggreen}{shell output}\tagchip{taggreen}{telemetry export}\tagchip{taggreen}{network message}} \\
    \multirow{2}{*}{Unauthorized actor} & \chips{\tagchip{tagorange}{UI viewer}\tagchip{tagorange}{other authenticated user}\tagchip{tagorange}{log reader}\tagchip{tagorange}{adjacent client}\tagchip{tagorange}{cross-tenant actor}\tagchip{tagorange}{infrastructure reader}} \\
     & \chips{\tagchip{tagorange}{chat recipient}\tagchip{tagorange}{workspace-restricted user}\tagchip{tagorange}{redirect target}\tagchip{tagorange}{SQL user}\tagchip{tagorange}{telemetry collector}} \\
    Repair pattern & \chips{\tagchip{tagpurple}{mask or redact}\tagchip{tagpurple}{add authorization check}\tagchip{tagpurple}{remove field}\tagchip{tagpurple}{defer resolution}\tagchip{tagpurple}{reject hardlink}\tagchip{tagpurple}{strip on redirect}\tagchip{tagpurple}{remove path echo}} \\
    \bottomrule
  \end{tabular}}
\end{table}

%% file: 4_BenchmarkStatistics.tex
\section{{\bench} Statistics}
\label{sec:statistics}

\begin{figure}[t]
  \centering
  \includegraphics[width=\linewidth]{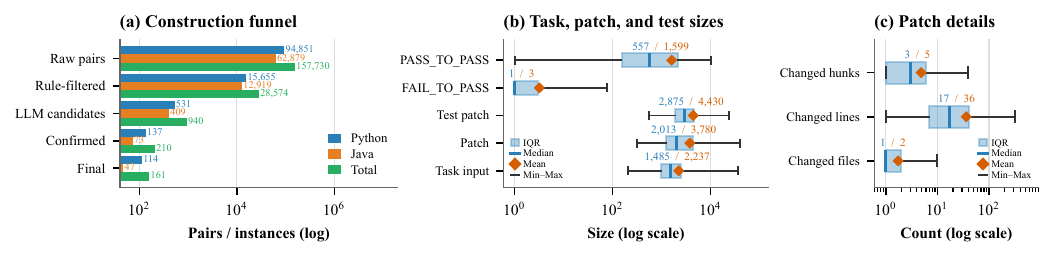}
  \caption{{\bench} overview: (a) construction funnel, (b) task, patch, and test statistics, and (c) patch details.}
  \label{fig:overview}
\end{figure}

\subsection{Benchmark Overview}
\label{subsec:benchmark-overview}

\subsubsection{Construction Funnel and Composition}
\label{subsubsec:funnel}

{\bench} comprises 161 task instances collected from 64 widely used open-source repositories, spanning both Python and Java projects. Specifically, the benchmark includes 114 Python instances and 47 Java instances, enabling evaluation of automated \infoex repair across both language ecosystems.

\begin{wrapfigure}{r}{0.32\linewidth}
  \centering
  \includegraphics[width=\linewidth]{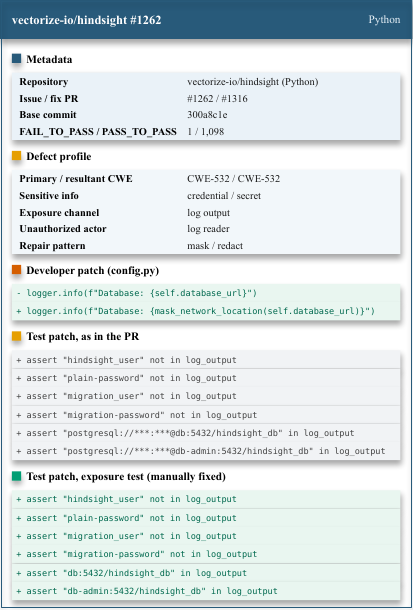}
  \caption{An example {\bench} instance.}
  \label{fig:example-instance}
\end{wrapfigure}
Figure~\ref{fig:overview}(a) summarizes the construction funnel. From 157,730 issue--PR pairs, rule-based filtering retained 28,574, LLM pre-screening 940, and human review confirmed 210 \infoex defects. Rule-based filtering keeps an issue--PR pair only when every changed file is source code, at least one changed file is a test, and at least one changed file is a non-test source file. Of the 129,156 discarded pairs, 87,777 (68.0\%) modify at least one non-source file, 36,566 (28.3\%) modify source code but no test, and 4,813 (3.7\%) modify only test files. Agreement among the three LLM assessments correlated with human confirmation: 177/313 (57\%) candidates flagged three times were confirmed, compared with 22/186 (12\%) flagged twice and 11/441 (2\%) flagged once. To assess false negatives from LLM pre-screening, we randomly sampled 600 pairs rejected by all three assessments. This sample size provides at least 95\% probability of detecting one or more missed defects if such defects constitute 0.5\% or more of the rejected pairs. Manual review using the same inclusion criteria identified one missed \infoex defect, corresponding to 0.17\% of the audited rejected candidates (95\% Wilson CI: 0.03\%--0.94\%). Execution verification retained 161 instances, roughly one per thousand initial pairs; the remaining 49 were excluded because tests depended on symbols introduced by the fix (24), behaved identically before and after the fix (23), or required unavailable external services (2). All subsequent analyses use these 161 instances. Figure~\ref{fig:example-instance} illustrates one retained instance, where a startup log exposes credentials from \texttt{DATABASE\_URL}; the developer patch masks them, and the exposure test checks that the username and password are absent. The figure also illustrates test normalization: in 57 instances, developer-written tests encode the exact reference-fix output, as in Figure~\ref{fig:example-instance}, where the test requires the literal redaction \texttt{***:***} and would reject alternative secure repairs, e.g., masking the same credentials with four asterisks instead of three. We therefore rewrite such assertions to check the exposure outcome rather than the specific implementation.

\subsubsection{Task, Patch, and Test Statistics}
\label{subsubsec:sizes}

\begin{wrapfigure}{r}{0.45\linewidth}
  \centering
  \includegraphics[width=\linewidth]{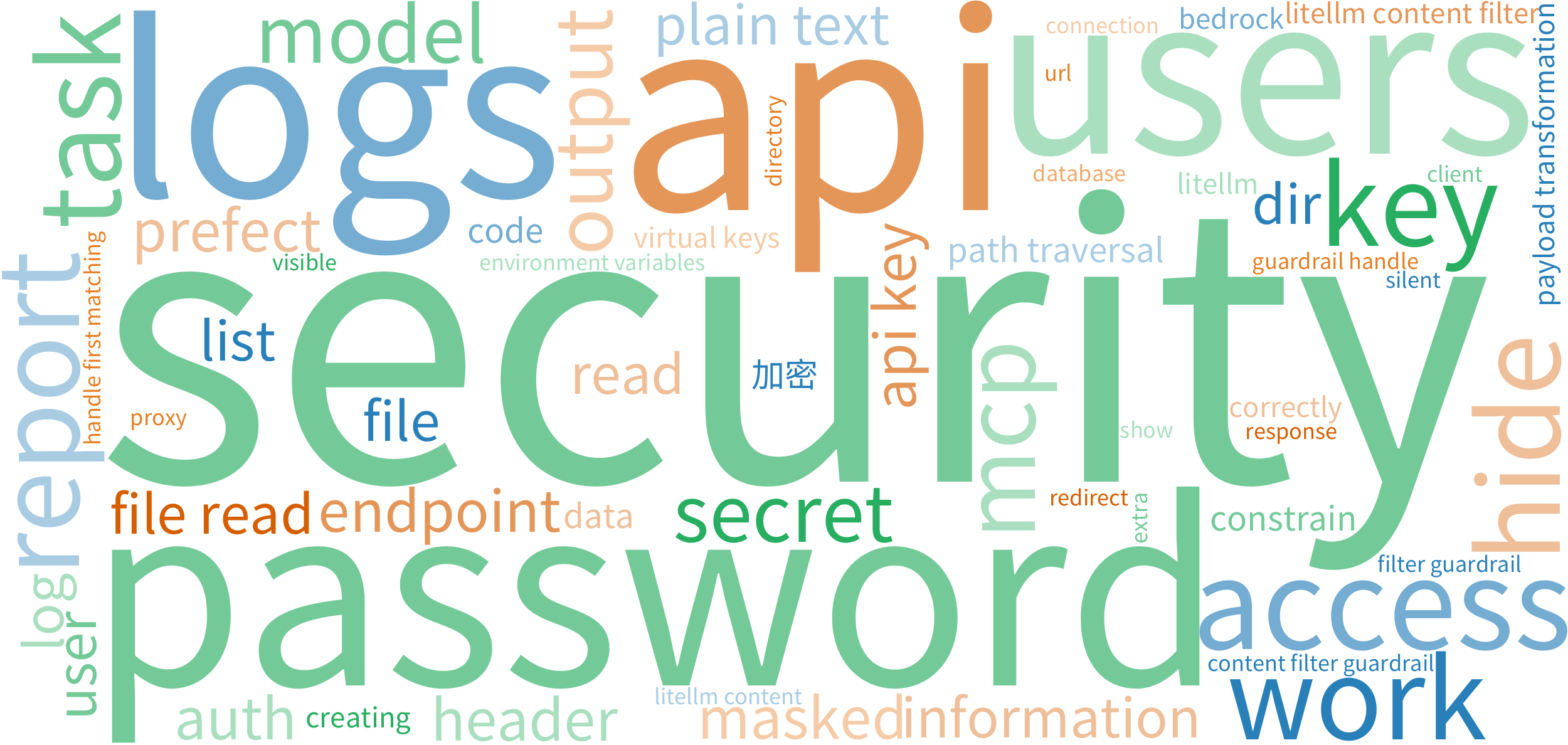}
  \caption{Issue-title keyphrases.}
  \label{fig:wordcloud}
\end{wrapfigure}
Figure~\ref{fig:overview}(b) summarizes task, patch, and test-suite sizes. Task descriptions, developer patches, and test patches have median lengths of 1,485, 2,013, and 2,875 characters, respectively. Each instance contains a median of one \texttt{FAIL\_TO\_PASS} test but 557 \texttt{PASS\_TO\_PASS} tests, reflecting the focused role of exposure tests and the broader role of regression tests. The distributions are right-skewed; for example, \texttt{PASS\_TO\_PASS} tests average 1,599 per instance. Figure~\ref{fig:overview}(c) shows that a typical developer patch changes one file, 17 lines, and three hunks, with two thirds of instances being single-file fixes, while larger multi-file changes provide variation in repair complexity. Finally, Figure~\ref{fig:wordcloud} highlights frequent issue-title keyphrases, including \emph{password}, \emph{privacy}, \emph{logs}, \emph{API}, \emph{user}, and \emph{access}, reflecting recurring themes in {\bench}.

\subsubsection{Temporal and Fix-Time Statistics}
\label{subsubsec:temporal}

\begin{wrapfigure}{r}{0.45\linewidth}
  \centering
  \includegraphics[width=\linewidth]{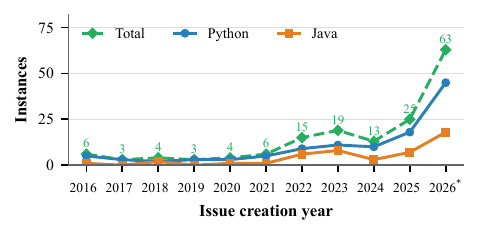}
  \caption{Instances by issue-creation year.}
  \label{fig:issueCreationYear}
\end{wrapfigure}
Figure~\ref{fig:issueCreationYear} shows the distribution of the 161 instances by issue-creation year, with 2026${}^{*}$ denoting a partial year covering issues before August 2026. {\bench} contains recent instances, with the annual count rising after 2022 and reaching 63 in partial-year 2026. Fix latency, measured from issue creation to the merge of the corresponding fix, has a median of 4 days and an interquartile range of 1--18 days. The distribution is strongly right-skewed, with a mean of 54 days and a maximum exceeding three years. Thus, {\bench} covers both rapidly resolved defects and substantially longer repair cycles.

\subsection{Source Distribution}
\label{subsec:source-distribution}

\subsubsection{Languages, Repositories, and Popularity}
\label{subsubsec:languages-repos}

The 161 instances come from 64 repositories, including 41 Python repositories with 114 instances and 23 Java repositories with 47 instances (Table~\ref{tab:repos}, Figure~\ref{fig:repo-stats}).
Repository popularity ranges from 7,200 to 89,500 GitHub stars, with a median of 27,600, while repository size ranges from 89 to 16,986 source files, with a median of 1,946. Instance counts are uneven: 40 repositories contribute one instance, whereas the six largest contributors, \texttt{keycloak}, \texttt{litellm}, \texttt{home-assistant}, \texttt{odysseus}, \texttt{prefect}, and \texttt{sentry}, account for 71 of the 161 instances.

\input{tables/repos}

\subsubsection{Application Domains}
\label{subsubsec:domains}

\begin{figure*}[t]
  \centering
  \includegraphics[width=\linewidth]{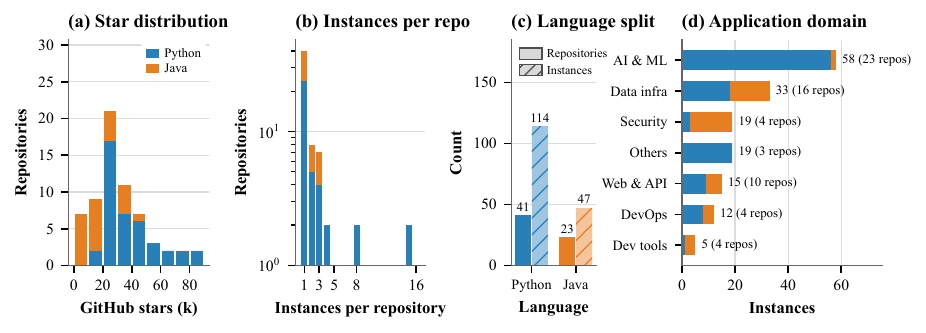}
  \caption{Repository statistics, including popularity, instance distribution, language coverage, and application domains. \emph{Others} groups three single-repository domains.}
  \label{fig:repo-stats}
\end{figure*}

We assign each repository a primary application domain based on its GitHub topics and description, manually review all assignments, and consolidate them into seven categories (Figure~\ref{fig:repo-stats}(d)). AI and machine learning is the largest domain, with 23 repositories and 58 instances, followed by data infrastructure with 16 repositories and 33 instances, and web and API frameworks with 10 repositories and 15 instances. Security and identity, DevOps and cloud, and developer tools each contribute four repositories, while three single-repository domains are grouped as \emph{Others}. These domains cover systems that routinely handle sensitive information such as credentials, prompts, user records, and access tokens. Notably, AI and machine-learning repositories first contribute instances in 2023 and account for 42 of the 63 instances created in 2026, reflecting their growing presence among recent \infoex defects in {\bench}. These repositories also exhibit domain-specific exposure patterns. For example, one LLM gateway masks one provider's key format but not another's when logging deployments, exposing an AWS access key and secret in plaintext. Such cases illustrate how modern AI software introduces heterogeneous credentials and providers that can create new opportunities for information exposure, underscoring the growing relevance of \infoex repair in this domain.

\subsection{Defect Distribution}
\label{subsec:defect-distribution}

\begin{figure*}[t]
  \centering
  \includegraphics[width=\linewidth]{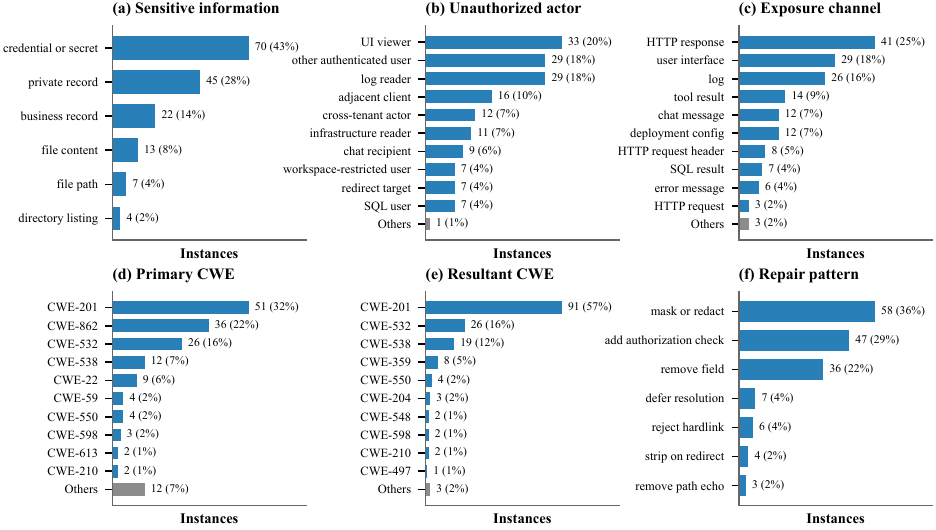}
  \caption{Distribution of the six characterization dimensions over 161 instances. The ten most frequent values are shown and the remainder grouped as \emph{Others}.}
  \label{fig:distribution}
\end{figure*}

Each instance is characterized along six dimensions: primary and resultant CWE, sensitive information, exposure channel, unauthorized actor, and repair pattern. Figure~\ref{fig:distribution} shows their distributions.

The exposure dimensions are dominated by several recurring categories. Credentials and secrets are the most common sensitive information (70 instances, 43.5\%), followed by private records (45, 28.0\%), together accounting for 71.4\% of {\bench}. The leading exposure channels are HTTP responses (41, 25.5\%), user interfaces (29, 18.0\%), and logs (26, 16.1\%), totaling 59.6\%. Unauthorized actors are similarly concentrated, with UI viewers, other authenticated users, and log readers each representing 18.0--20.5\% of the benchmark.

The CWE dimensions further show that the weaknesses causing an exposure often differ from the weaknesses through which the exposure ultimately manifests. By definition, all resultant CWEs in {\bench} fall within the CWE-200 hierarchy, and they are concentrated in a small number of categories, particularly CWE-201 (sensitive information in sent data; 56.5\%), CWE-532 (sensitive information in log files; 16.1\%), and CWE-538 (sensitive information in externally accessible files or directories; 11.8\%). In contrast, primary CWEs are more diverse and are not restricted to this hierarchy. CWE-201 serves as the primary CWE in 31.7\% of instances, while CWE-862 (missing authorization; 22.4\%) and CWE-22 (path traversal; 5.6\%) are notable primary root causes that lie outside the CWE-200 hierarchy. Overall, the primary and resultant CWEs are identical in 63.4\% of instances, whereas 35.4\% of instances originate from primary weaknesses outside the CWE-200 hierarchy. %
Consequently, restricting benchmark construction to defects whose root causes fall within CWE-200 would exclude more than one third of the observed \infoex defects.

%% file: tables/repos.tex
\begin{table}[t]
  \centering
  \caption{Repositories contributing to {\bench}, sorted by instance count. The top 20 are shown individually and the remaining 44 are aggregated.}
  \label{tab:repos}
  \resizebox{\linewidth}{!}{%
  \begin{tabular}{@{}l c l r r r r r l l@{}}
    \toprule
    \textbf{Repository} & \textbf{Lang} & \textbf{License} & \textbf{Stars} &
    \textbf{\#Files} & \textbf{Size (MB)} & \textbf{\#LOC} & \textbf{\#Inst} &
    \textbf{Domain} & \textbf{Description} \\
    \midrule
    \texttt{keycloak/keycloak} & Java & Apache-2.0 & 36k & 9,364 & 589 & 1,320,860 & 16 & Security \& Identity & Identity and access management server for modern applications and services \\
    \texttt{BerriAI/litellm} & Py & NOASSERTION & 55k & 3,998 & 1243 & 1,041,068 & 15 & AI \& Machine Learning & Gateway and SDK for calling many LLM APIs \\
    \texttt{home-assistant/core} & Py & Apache-2.0 & 89k & 9,518 & 838 & 1,617,741 & 15 & Home Automation & Open source home automation platform prioritizing local control \\
    \texttt{odysseus-dev/odysseus} & Py & AGPL-3.0 & 62k & 913 & 42 & 331,091 & 9 & AI \& Machine Learning & Self-hosted AI workspace for tools and workflows \\
    \texttt{PrefectHQ/prefect} & Py & Apache-2.0 & 23k & 2,139 & 224 & 487,610 & 8 & Data Infrastructure & Workflow orchestration framework for building resilient data pipelines \\
    \texttt{getsentry/sentry} & Py & NOASSERTION & 44k & 6,254 & 896 & 1,041,806 & 8 & DevOps \& Cloud & Developer-first error tracking and performance monitoring platform \\
    \texttt{HKUDS/nanobot} & Py & MIT & 46k & 487 & 51 & 164,127 & 5 & AI \& Machine Learning & Lightweight open source AI agent for tools and chats \\
    \texttt{agno-agi/agno} & Py & Apache-2.0 & 41k & 4,838 & 294 & 846,146 & 4 & AI \& Machine Learning & Framework for building and managing AI agent platforms \\
    \texttt{apache/airflow} & Py & Apache-2.0 & 46k & 7,172 & 661 & 1,214,695 & 4 & Data Infrastructure & Platform to author schedule and monitor data workflows \\
    \texttt{PrefectHQ/fastmcp} & Py & Apache-2.0 & 27k & 628 & 55 & 144,123 & 3 & AI \& Machine Learning & Python framework for building MCP servers and clients \\
    \texttt{StarRocks/starrocks} & Java & Apache-2.0 & 12k & 7,101 & 732 & 1,537,497 & 3 & Data Infrastructure & Distributed OLAP query engine for real-time analytics \\
    \texttt{halo-dev/halo} & Java & GPL-3.0 & 39k & 1,025 & 86 & 119,300 & 3 & Web \& API & Open source CMS and website building platform \\
    \texttt{kestra-io/kestra} & Java & Apache-2.0 & 28k & 2,775 & 121 & 264,901 & 3 & Data Infrastructure & Event driven orchestration and scheduling platform for workflows \\
    \texttt{openai/openai-agents-python} & Py & MIT & 28k & 1,260 & 33 & 302,045 & 3 & AI \& Machine Learning & Lightweight framework for building multi-agent workflows \\
    \texttt{paperless-ngx/paperless-ngx} & Py & GPL-3.0 & 44k & 650 & 190 & 84,029 & 3 & Document Management & Document management system to scan index and archive documents \\
    \texttt{zulip/zulip} & Py & Apache-2.0 & 26k & 2,695 & 622 & 504,163 & 3 & Web \& API & Open source team chat server and web application \\
    \texttt{PostHog/posthog} & Py & NOASSERTION & 37k & 6,351 & 5700 & 1,203,896 & 2 & Data Infrastructure & Product analytics platform with session replay and feature flags \\
    \texttt{apache/shardingsphere} & Java & Apache-2.0 & 21k & 9,839 & 684 & 715,560 & 2 & Data Infrastructure & Distributed SQL database middleware for sharding and data encryption \\
    \texttt{apache/superset} & Py & Apache-2.0 & 74k & 4,848 & 1046 & 925,660 & 2 & Data Infrastructure & Open-source business intelligence and data visualization web platform \\
    \texttt{deepset-ai/haystack} & Py & Apache-2.0 & 26k & 436 & 211 & 99,044 & 2 & AI \& Machine Learning & Framework for building LLM and retrieval-augmented generation applications \\
    \midrule
    \textit{Remaining 44 repositories} & -- & -- & -- & 132,562 & 9,725 & 21,057,583 & 48 & \multicolumn{2}{l}{Omitted for brevity} \\
    \midrule
    \textbf{Total} & & & & 214,853 & 24,043 & 35,022,945 & \textbf{161} & \multicolumn{2}{l}{64 repositories} \\
    \bottomrule
  \end{tabular}}
\end{table}

%% file: 5_ExperimentalSetup.tex
\section{Experimental Setup}
\label{sec:setup}

\subsection{Research Questions}
\label{subsec:rq}

\newcommand{\firstRQ}{How do current repair methods perform on {\bench}?}
\newcommand{\shortFirstRQ}{Overall Performance}

\newcommand{\secondRQ}{How does repair performance vary across defect characteristics?}
\newcommand{\shortSecondRQ}{Stratified Performance}

\newcommand{\thirdRQ}{Which failure modes limit automated repair on {\bench}?}
\newcommand{\shortThirdRQ}{Failure Analysis}

\newcommand{\fourthRQ}{How does inference budget affect repair performance?}
\newcommand{\shortFourthRQ}{Budget Sensitivity}

We investigate the following research questions:
\begin{enumerate}[leftmargin=*, label=\textbf{RQ\arabic*}]
    \item {\firstRQ} (\textbf{\shortFirstRQ})
    \item {\secondRQ} (\textbf{\shortSecondRQ})
    \item {\thirdRQ} (\textbf{\shortThirdRQ})
\end{enumerate}

\subsection{Studied Systems}
\label{subsec:studied-systems}

We evaluate six systems in two groups. The first includes issue-resolution methods commonly used in prior research, spanning procedural and agentic designs. \textbf{MSWE-agent~\cite{githubGitHubMultiswebenchMSWEagent}}, \textbf{Magentless~\cite{githubGitHubMultiswebenchMagentless}}, and \textbf{MopenHands~\cite{githubGitHubMultiswebenchMopenhands}} are the Multi-SWE-bench~\cite{zan2026multi} multilingual forks of SWE-agent~\cite{yang2024sweAgent}, Agentless~\cite{xia2025agentless}, and OpenHands~\cite{wang2024openhands}, enabling the same methods to cover both Python and Java. We also include \textbf{mini-swe-agent~\cite{githubGitHubSWEagentminisweagent}}, a minimal bash-only agent widely used as a reference baseline~\cite{xia2025livesweagent,huang2026deepswemeasuringfrontiercoding,deng2025swebenchproaiagents}. The second group includes developer-facing coding agents: the open-source \textbf{OpenCode~\cite{githubGitHubAnomalycoopencode}} and commercial \textbf{Codex~\cite{githubGitHubOpenaicodex}}.

All agentic systems receive a budget of 100 interaction steps, while Magentless uses a $10 \times 10$ patch space comprising 10 localization groups and 10 sampled patches per group.
To prevent unintended access to external information or future repository history, we configure the Docker containers to block all network access except LLM API requests and remove all commits after the buggy commit.

\subsection{Model Selection}
\label{subsec:model-selection}

We conduct evaluation experiments with two groups of base LLMs.

\input{tables/models}

As shown in Table~\ref{tab:models}, the first group includes \textbf{DeepSeek-V3.2~\cite{deepseekai2025deepseekv32pushingfrontieropen}}, \textbf{Qwen3.5-Plus~\cite{qwen3.5}}, and \textbf{GPT-5.6 Luna~\cite{gpt56luna}}, crossed with all six studied systems, yielding $3 \times 6$ configuration pairs. We select these models to balance capability, cost, and release date, covering competitive open-weight models and a cost-effective proprietary model. To support leakage-aware analysis, we split the 161 {\bench} instances at 2026-02-16 (the latest knowledge cutoff among the three models): 105 have issue dates on or before 2026-02-16 and 56 after it. DeepSeek-V3.2 and Qwen3.5-Plus were released before the boundary, while GPT-5.6 Luna has a documented knowledge cutoff of 2026-02-16. Thus, none of the three models could have had access to the 56 post-boundary instances during training.

The second group comprises more recent models released after the boundary: \textbf{DeepSeek-V4.1-Flash~\cite{deepseekai2026deepseekv41flashpushinglimitskv}}, \textbf{GLM-5.3-Flash~\cite{glm53flash}}, \textbf{Qwen3.8-Flash~\cite{qwen38flash}} and \textbf{GPT-6 Luna~\cite{gpt6luna}}. Following prior benchmark studies~\cite{deng2025swebenchproaiagents,huang2026deepswemeasuringfrontiercoding}, we evaluate these models only with mini-swe-agent to isolate model effects under a fixed agent. As these models postdate the boundary, their results are reported separately from those of the first group.

\subsection{Evaluation Metrics}
\label{subsec:metrics}

Following prior work \cite{jimenez2023swebench, yang2024sweAgent, xia2025agentless, wang2024openhands, guo2025omnigirl}, we report the following metrics:
\begin{itemize}
    \item \textbf{\#Resolved (\%Resolved)}: the number (and the corresponding percentage) of
    benchmark instances that are successfully resolved; the tables abbreviate the metric as
    \#R (\%R).
    \item \textbf{Avg. \$}: the average cost per instance in USD.
\end{itemize}

%% file: tables/models.tex
\begin{table}[H]
\centering
\small
\caption{Base LLMs in the first (leakage-controlled) model group.}
\label{tab:models}
\begin{tabular}{lllcc}
\toprule
\textbf{Model} & \textbf{Open} & \textbf{Parameters} & \textbf{Release} & \textbf{Knowledge cutoff} \\
\midrule
DeepSeek-V3.2 & Yes & 671B (37B active) & 2025-12-01 & Not published \\
Qwen3.5-Plus & Yes & 397B (17B active) & 2026-02-16 & Not published \\
GPT-5.6 Luna & No & Undisclosed & 2026-07-09 & 2026-02-16\footnotemark \\
\bottomrule
\end{tabular}
\end{table}
\footnotetext{\url{https://developers.openai.com/api/docs/models/gpt-5.6-luna}}

%% file: 6_Evaluation.tex
\input{tables/rq1_group12}

\section{Evaluation}
\label{sec:evaluation}

\subsection{RQ1: {\shortFirstRQ}}
\label{subsec:rq1}

We first evaluate six repair methods with three LLMs, yielding 18 configurations over all 161 instances. Table~\ref{tab:rq1group12} reports repair success, test outcomes, API cost, and interaction steps for each configuration. Here, \emph{F2P} denotes the normalized \texttt{FAIL\_TO\_PASS} tests for exposure removal, \emph{$F2P'$} the original developer-written tests before normalization, and \emph{P2P} the \texttt{PASS\_TO\_PASS} regression tests; an instance is resolved only if both F2P and P2P pass. %

\textbf{Overall repair effectiveness.}
Resolve rates range from 9.9\% for Magentless with Qwen3.5-Plus to 39.1\% for MopenHands with the same LLM (63/161), with a mean of 30.3\% across all 18 configurations. Thus, substantial room remains even for the strongest evaluated configuration.

P2P pass rates are consistently higher than exposure-test pass rates, ranging from 32.3\% to 81.4\% (65.0\% on average), compared with 12.4\%--47.8\% (36.0\%) for F2P and 9.9\%--37.3\% (29.1\%) for F2P$'$. This indicates that eliminating the exposure is more challenging than preserving previously passing behavior under our protocol. Test normalization also matters: at least one configuration passes the normalized exposure tests for 103 instances, compared with 89 under the original developer-written tests, reducing rejections caused by implementation-specific assertions.

\textbf{Method and model effects.}
Averaged over methods, the three LLMs achieve similar resolve rates of 30.2\%--30.3\%, whereas method-level averages range from 14.5\% for Magentless to 35.0\% for Codex. Thus, among the similarly capable models evaluated here, performance varies more across repair methods than across LLMs. Resource use also varies widely, with average cost ranging from \$0.007 to \$0.401 per instance and \$0.019 to \$2.052 per resolved instance.

\begin{wrapfigure}{r}{0.3\linewidth}
  \centering
  \includegraphics[width=\linewidth]{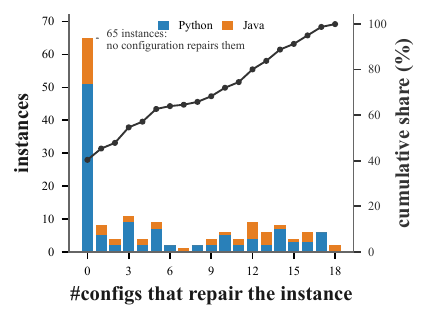}
  \caption{Distribution of instances by the number of configurations that resolve them.}
  \label{fig:rq1coverage}
\end{wrapfigure}

Java instances outperform Python instances in 17 of 18 configurations, averaging 37.0\% and 27.5\%, respectively. Repository composition, however, precludes attributing this gap to language alone. Resolve rates are similar before and after the leakage boundary (29.5\% vs.\ 31.6\%), providing no aggregate evidence of a pre-boundary advantage while not ruling out contamination of individual instances.

\textbf{Instance coverage.}
Figure~\ref{fig:rq1coverage} shows that 65 of 161 instances (40.4\%) remain unresolved by all 18 configurations. Among the 96 resolved by at least one configuration, 51 are resolved by ten or more. The union of all configurations therefore reaches 59.6\% coverage, compared with 39.1\% for the best individual configuration. This bimodal pattern suggests that many instances are either broadly solvable or consistently difficult.

\begin{wrapfigure}{r}{0.3\linewidth}
  \centering
  \includegraphics[width=\linewidth]{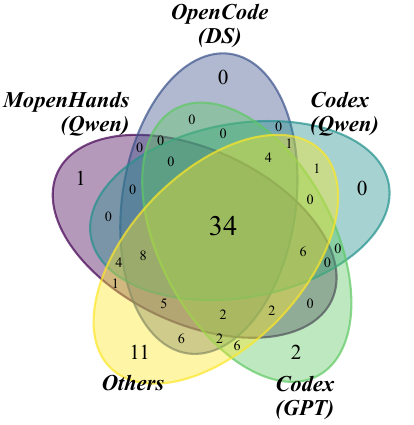}
  \caption{Repair overlap across configurations.}
  \label{fig:rq1vennD}
\end{wrapfigure}

\textbf{Complementarity across configurations.}
No single configuration subsumes the repairs of the others. As shown in Figure~\ref{fig:rq1vennD}, the four highest-performing configurations collectively resolve 85 instances, 22 more than the best configuration, while the remaining 14 contribute 11 additional repairs, bringing total coverage to 96 instances (59.6\%). Thus, configurations exhibit complementary repair capabilities, although 40.4\% of {\bench} remains unresolved.

\input{tables/rq1_group3}

\textbf{Model sensitivity under a fixed agent.}
To isolate LLM choice from the repair workflow, we fix the method to mini-swe-agent and compare additional models. Among the three primary LLMs, DeepSeek-V3.2 and GPT-5.6 Luna each resolve 56 instances (34.8\%), while Qwen3.5-Plus resolves 54 (33.5\%). Their resolve rates are similar, but costs vary substantially, with GPT-5.6 Luna being the cheapest (\$0.007 per instance and \$0.019 per resolved instance). \textbf{We further evaluate four newer LLMs:} DeepSeek-V4.1-Flash, GLM-5.3-Flash, Qwen3.8-Flash, and GPT-6 Luna. Because they postdate the leakage boundary, we report them separately from the primary 18 configurations (Table~\ref{tab:rq1group3}). All four improve under the same agent, resolving 58--65 instances (36.0\%--40.4\%). DeepSeek-V4.1-Flash and GLM-5.3-Flash each resolve 65 (40.4\%), exceeding the best primary configuration at 63 (39.1\%), although these results are not leakage-controlled. The newer models are also cheaper (\$0.012--\$0.082 per resolved instance) and generally require fewer steps than the two step-heavy primary models.

\begin{rqsummary}
\textbf{RQ1 Summary:}
Current systems remain limited on {\bench}: the best of the 18 configurations resolves 39.1\% of instances, while 40.4\% remain unresolved by all. Exposure removal is harder than preserving existing behavior, and test normalization prevents implementation-specific assertions from rejecting secure repairs. Complementary configurations increase overall coverage to 59.6\%. Under a fixed agent, newer post-boundary models achieve 36.0\%--40.4\% resolve rates, though these results are not leakage-controlled.
\end{rqsummary}

\input{tables/rq2}

\subsection{RQ2: {\shortSecondRQ}}
\label{subsec:rq2}

Whereas RQ1 examines aggregate repair performance, RQ2 investigates how effectiveness varies across defect characteristics. We stratify the 161 instances along eight characteristics, comprising the six characterization dimensions, primary-CWE scope, and application domain. The experimental results are shown in Table~\ref{tab:rq2}.

\textbf{Repairability varies across defect characteristics.}
Resolve rates differ markedly across both characteristic values and configurations. For example, among credential and secret exposures, individual configurations resolve 14.3\%--44.3\% of instances. Moreover, no configuration achieves the highest resolve rate in more than 16 of the 51 fine-grained categories. Thus, aggregate performance alone does not capture how repair effectiveness varies across different forms of \infoex defects.

\textbf{Direct exposure mitigation is more tractable than structural repair.}
A consistent pattern emerges across repair patterns and root causes. Instances repaired through masking or redaction achieve a mean resolve rate of 37.5\%, compared with 26.5\% for authorization checks and 17.5\% for deferred resolution. Similarly, CWE-532 defects involving sensitive information in logs average 43.2\%, whereas CWE-862 defects involving missing authorization average 22.7\%. These results suggest that current systems more readily handle defects where the exposed value can be directly suppressed than those requiring changes to authorization or other underlying program behavior.

\textbf{Root causes outside CWE-200 are harder on average.}
Instances whose primary CWE lies outside the CWE-200 hierarchy achieve a lower mean resolve rate than those within it (25.1\% vs.\ 33.1\%). Nevertheless, 57.9\% of these instances are resolved by at least one configuration, compared with 35.1\% for the best individual configuration. Thus, these defects are challenging for individual systems rather than uniformly unsolvable. This finding also supports including resultant exposures whose root causes fall outside CWE-200, as restricting the benchmark to the hierarchy would omit a substantial and relatively difficult subset.

\begin{rqsummary}
\textbf{RQ2 Summary:}
Repairability varies markedly across defect characteristics, with no configuration consistently leading across categories. Defects addressed through direct sanitization are repaired more successfully than those requiring authorization or other structural changes. Moreover, exposures originating outside the CWE-200 hierarchy are harder on average, underscoring the importance of including such resultant exposures in repair evaluation.
\end{rqsummary}

\input{tables/rq3_failure_causes}

\subsection{RQ3: {\shortThirdRQ}}
\label{subsec:rq3}

\begin{figure}[tbp]
  \centering
  \includegraphics[width=\linewidth]{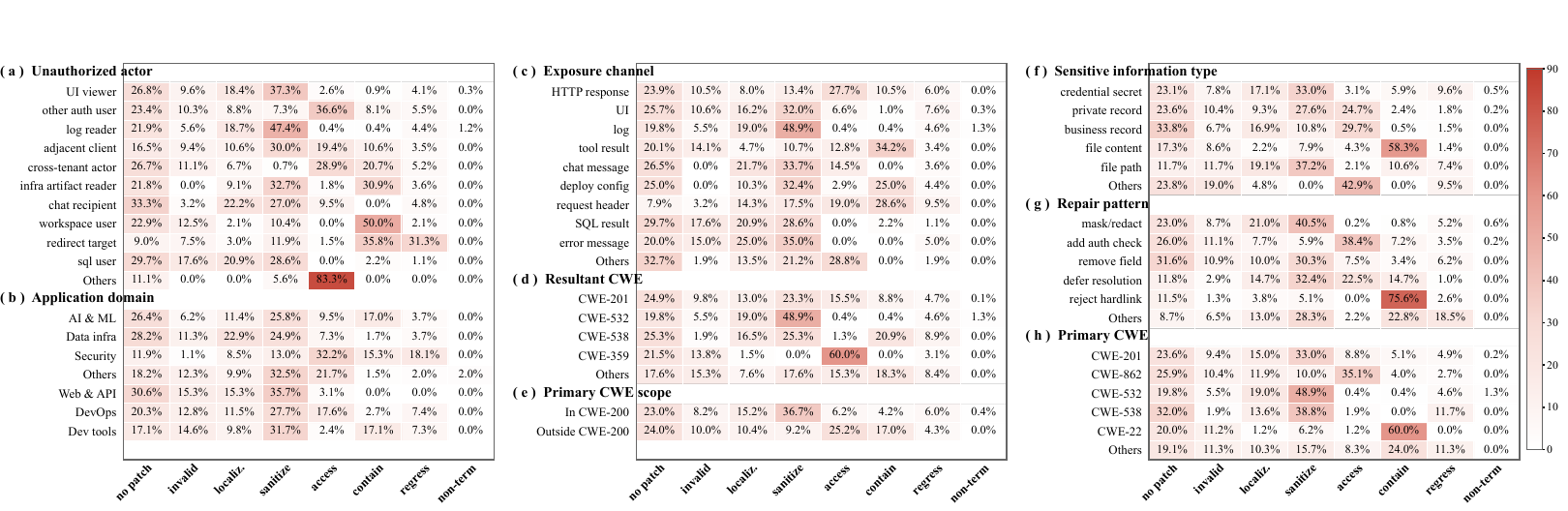}
  \caption{Failure-family distributions across the eight defect characteristics in Table~\ref{tab:rq2}, row-normalized by characteristic value. The eight families are no patch (C1), invalid patch (C2), localization (C3.1), sanitization (C3.2--C3.6), access control and state (C3.7--C3.8), containment and validation (C3.9--C3.10), regression (C4.1--C4.3), and non-termination (C5).}
  \label{fig:rq3causechar}
\end{figure}

\begin{figure}[tbp]
  \centering
  \includegraphics[width=\linewidth]{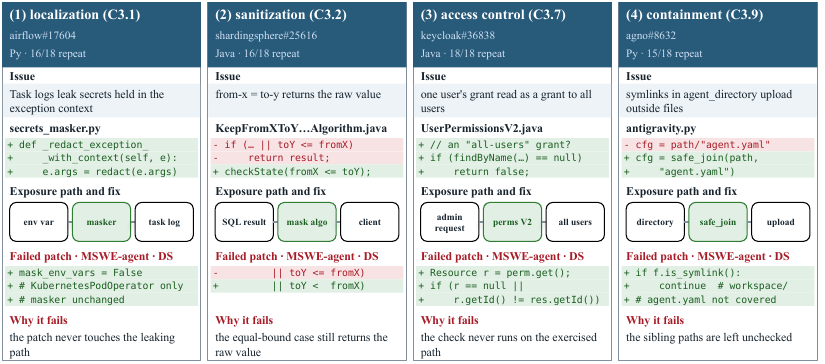}
  \caption{Four representative cases illustrating distinct failure families on instances unresolved by all 18 configurations: localization (C3.1), sanitization (represented by C3.2), access control and state (represented by C3.7), and containment and validation (represented by C3.9).}
  \label{fig:rq3cases}
\end{figure}

To investigate the limitations identified in RQ1 and RQ2, we analyze the unsuccessful configuration--instance attempts across all 18 configurations. We classify each failure according to its primary cause (Table~\ref{tab:rq3causes}), examine how these causes vary with defect characteristics (Figure~\ref{fig:rq3causechar}), and illustrate recurring failure mechanisms through four representative cases (Figure~\ref{fig:rq3cases}). Our taxonomy distinguishes five main outcomes: no source change (C1), an invalid patch (C2), a valid patch that leaves the exposure unresolved (C3), a patch that eliminates the exposure but introduces functional regressions (C4), and non-termination (C5). We further decompose C3 and C4 into specific causes. All percentages are computed over failed attempts, with each attempt assigned to exactly one main cause and, where applicable, one corresponding subcategory.

\textbf{Incomplete exposure removal dominates failures.}
As shown in Table~\ref{tab:rq3causes}, most failed attempts produce a valid patch that still leaves the exposure unresolved (C3, 62.1\%). By comparison, 23.4\% produce no source change (C1), 8.9\% produce an invalid patch (C2), 5.4\% remove the exposure but introduce regressions (C4), and 0.3\% do not terminate (C5). Thus, the main challenge is not merely generating a valid patch or preserving functionality, but actually eliminating the exposure.

A closer examination of C3 shows that sanitization and information-handling failures (C3.2--C3.6) account for 26.0\% of all failures, including 12.0\% from incomplete or reversible sanitization alone. Access-control and state-management failures (C3.7--C3.8) contribute another 13.5\%, followed by localization errors (C3.1, 13.3\%) and containment or request-validation failures (C3.9--C3.10, 9.2\%). Although localization is the largest individual subcategory, it accounts for only about one fifth of C3 failures. Nearly four fifths arise after systems reach the relevant code, indicating that localization alone is insufficient. Successful repair must also account for the conditions, representations, and execution paths through which sensitive information remains accessible.

This pattern is even stronger for the 65 instances unresolved by every configuration. C3 accounts for 71.5\% of their failed attempts, compared with 53.0\% for instances resolved by at least one configuration. Thus, consistently unresolved instances are characterized less by a failure to generate patches than by repeated failure to eliminate the underlying exposure.

\textbf{Failure profiles differ across configurations.}
Although C3 dominates overall, failure profiles vary substantially across configurations. No-patch failures (C1) account for 3.9\%--59.2\% of unsuccessful attempts, invalid patches (C2) for 0\%--22.0\%, and surviving exposures (C3) for 27.5\%--89.6\%. For example, Magentless has more no-patch failures than the agentic methods (36.8\% vs.\ 19.5\%) but fewer surviving exposures (49.3\% vs.\ 65.8\%). Thus, Magentless more often fails to produce a patch, whereas agentic methods more often generate executable patches that leave the exposure unresolved. Overall resolve rates therefore obscure where different repair processes fail.

\textbf{Failure mechanisms are associated with defect characteristics.}
Figure~\ref{fig:rq3causechar} further relates the failure taxonomy to the characteristics examined in RQ2. Distinct exposure mechanisms exhibit markedly different failure profiles. Sanitization failures dominate log exposures (48.9\%), whereas containment and validation failures are more common for tool-result exposures (34.2\%) and especially file-content exposures (58.3\%). These results indicate that unsuccessful repairs fail in different ways depending on how sensitive information becomes accessible.

The CWE dimensions provide consistent evidence. Sanitization accounts for 48.9\% of failures involving CWE-532 (Insertion of Sensitive Information into Log File), while access-control and state failures account for 35.1\% of those rooted in CWE-862 (Missing Authorization). For CWE-22 (Path Traversal), containment and validation dominate at 60.0\%. Similar associations appear in repair patterns: 40.5\% of failed masking or redaction repairs involve sanitization problems, whereas 38.4\% of failed authorization-check repairs involve access-control and state failures. Defects requiring hard-link rejection exhibit an even stronger concentration in containment-related failures (75.6\%).

Taken together, these results show that unsuccessful repairs are not uniformly distributed across failure mechanisms. Rather, their composition is closely associated with the exposure mechanism, underlying weakness, and required mitigation. Thus, different defect types differ not only in repair success, as shown in RQ2, but also in the technical failure modes encountered during repair.

\textbf{Case studies of recurring incomplete repairs.}
To illustrate how incomplete repairs arise in practice, Figure~\ref{fig:rq3cases} presents four instances unresolved by all 18 configurations. Each represents a distinct C3 failure family and contrasts the developer patch and exposure path with an unsuccessful patch from MSWE-agent with DeepSeek-V3.2.

Airflow \#17604 (C3.1) illustrates localization failure: the developer patch sanitizes exception arguments exposing retained secrets through task logs, whereas the failed patch modifies environment-variable masking but misses the exposure path. ShardingSphere \#25616 (C3.2) illustrates incomplete sanitization: the failed patch changes a boundary condition but allows the equality case to return an unsanitized value. In both cases, generated changes modify related code but do not block the exposure path (Airflow \#17604) or block it only partially (ShardingSphere \#25616).

The other two cases concern authorization and path containment. In Keycloak \#36838 (C3.7), the generated patch adds a resource-identity check that is not reached on the path causing unauthorized access, while the developer fix corrects the underlying permission evaluation. In Agno \#8632 (C3.9), the generated patch rejects some symbolic links but leaves sibling upload paths insufficiently constrained, whereas the developer fix applies safe path resolution to prevent uploads outside the intended directory. Again, the generated changes modify security-related code without blocking the exposure: the added identity check never runs on the exercised path (Keycloak \#36838), and the symlink rejection leaves the sibling paths unchecked (Agno \#8632).

Across these four instances, 15--18 of the 18 configurations exhibit the corresponding failure. Their recurrence across methods and LLMs points to a common limitation: generated patches often address a visible manifestation of the defect without closing all exposure paths. This motivates future work that explicitly discovers and validates all relevant exposure paths, ensuring that a repair closes all of them rather than only the one it edits.

\begin{rqsummary}
\textbf{RQ3 Summary:}
Incomplete exposure removal dominates failures, accounting for 62.1\% of unsuccessful attempts. Nearly four fifths of these failures occur after the relevant code is reached, driven mainly by sanitization, access-control, and containment failures rather than localization alone. Representative cases further show that systems often modify related code without fully closing the exposure path. Effective \infoex repair therefore requires discovering and eliminating all relevant exposure paths.
\end{rqsummary}

%% file: tables/rq1_group12.tex
\definecolor{rqgreen}{HTML}{27AE60}
\begin{table}[t]
  \centering
  \small
  \caption{Overall repair performance of the 18 configurations. Column-best values are bolded and darkest-shaded, with lower cost preferred; steps are reported as min--max (mean).}
  \label{tab:rq1group12}
  \setlength{\tabcolsep}{9pt}
  \resizebox{\linewidth}{!}{%
  \begin{tabular}{@{}c l l c r r r r r r r r r r@{}}
    \toprule
    \multirow{2}{*}[-2.7pt]{\normalsize\textbf{Type}} & \multirow{2}{*}[-2.7pt]{\normalsize\textbf{Method}} & \multirow{2}{*}[-2.7pt]{\normalsize\textbf{Model}} & \multirow{2}{*}[-2.7pt]{\normalsize\textbf{Steps}} & \multicolumn{2}{c}{\textbf{\#R (\%R) on leakage subsets}} & \multicolumn{2}{c}{\textbf{\#R (\%R) by language}} & \multicolumn{3}{c}{\textbf{Pass details per test suite}} & \multirow{2}{*}[-2.7pt]{\normalsize\textbf{\#R (\%R)}} & \multirow{2}{*}[-2.7pt]{\normalsize\textbf{Avg.\ \$}} & \multirow{2}{*}[-2.7pt]{\normalsize\textbf{\$/Solved}} \\
    \cmidrule(lr){5-6} \cmidrule(lr){7-8} \cmidrule(lr){9-11}
    & & & & \textbf{$\le$02-16 (105)} & \textbf{$>$02-16 (56)} & \textbf{Python (114)} & \textbf{Java (47)} & \textbf{\#F2P (\%F2P)} & \textbf{\#F2P$'$ (\%F2P$'$)} & \textbf{\#P2P (\%P2P)} & & & \\
    \midrule
    \multirow{12}{*}{\rotatebox{90}{Issue-resolution systems}} & \multirow{3}{*}{MSWE-agent~\cite{githubGitHubMultiswebenchMSWEagent}} & DeepSeek-V3.2 & \makebox[1.15em][r]{15}--\makebox[1.75em][l]{100}~(72) & \cellcolor{rqgreen!46}27 (25.7\%) & \cellcolor{rqgreen!46}15 (26.8\%) & \cellcolor{rqgreen!36}22 (19.3\%) & \cellcolor{rqgreen!66}20 (42.6\%) & \cellcolor{rqgreen!46}50 (31.1\%) & \cellcolor{rqgreen!46}39 (24.2\%) & \cellcolor{rqgreen!40}87 (54.0\%) & \cellcolor{rqgreen!48}42 (26.1\%) & \$0.119 & \$0.456 \\
     &  & Qwen3.5-Plus & \makebox[1.15em][r]{7}--\makebox[1.75em][l]{100}~(57) & \cellcolor{rqgreen!49}28 (26.7\%) & \cellcolor{rqgreen!59}18 (32.1\%) & \cellcolor{rqgreen!50}29 (25.4\%) & \cellcolor{rqgreen!53}17 (36.2\%) & \cellcolor{rqgreen!53}56 (34.8\%) & \cellcolor{rqgreen!57}46 (28.6\%) & \cellcolor{rqgreen!50}98 (60.9\%) & \cellcolor{rqgreen!54}46 (28.6\%) & \$0.401 & \$1.405 \\
     &  & GPT-5.6 Luna & \makebox[1.15em][r]{8}--\makebox[1.75em][l]{100}~(32) & \cellcolor{rqgreen!53}30 (28.6\%) & \cellcolor{rqgreen!72}21 (37.5\%) & \cellcolor{rqgreen!52}30 (26.3\%) & \cellcolor{rqgreen!71}21 (44.7\%) & \cellcolor{rqgreen!59}60 (37.3\%) & \cellcolor{rqgreen!60}48 (29.8\%) & \cellcolor{rqgreen!58}107 (66.5\%) & \cellcolor{rqgreen!62}51 (31.7\%) & \$0.065 & \$0.204 \\
    \cmidrule(lr){2-14}
     & \multirow{3}{*}{Magentless~\cite{githubGitHubMultiswebenchMagentless}} & DeepSeek-V3.2 & \makebox[1.15em][r]{}--\makebox[1.75em][l]{} & \cellcolor{rqgreen!19}15 (14.3\%) & \cellcolor{rqgreen!25}10 (17.9\%) & \cellcolor{rqgreen!28}18 (15.8\%) & \cellcolor{rqgreen!8}7 (14.9\%) & \cellcolor{rqgreen!19}29 (18.0\%) & \cellcolor{rqgreen!21}24 (14.9\%) & \cellcolor{rqgreen!43}90 (55.9\%) & \cellcolor{rqgreen!22}25 (15.5\%) & \$0.104 & \$0.671 \\
     &  & Qwen3.5-Plus & \makebox[1.15em][r]{}--\makebox[1.75em][l]{} & \cellcolor{rqgreen!8}10 (9.5\%) & \cellcolor{rqgreen!8}6 (10.7\%) & \cellcolor{rqgreen!8}8 (7.0\%) & \cellcolor{rqgreen!12}8 (17.0\%) & \cellcolor{rqgreen!8}20 (12.4\%) & \cellcolor{rqgreen!8}16 (9.9\%) & \cellcolor{rqgreen!8}52 (32.3\%) & \cellcolor{rqgreen!8}16 (9.9\%) & \$0.204 & \$2.052 \\
     &  & GPT-5.6 Luna & \makebox[1.15em][r]{}--\makebox[1.75em][l]{} & \cellcolor{rqgreen!26}18 (17.1\%) & \cellcolor{rqgreen!29}11 (19.6\%) & \cellcolor{rqgreen!30}19 (16.7\%) & \cellcolor{rqgreen!22}10 (21.3\%) & \cellcolor{rqgreen!27}35 (21.7\%) & \cellcolor{rqgreen!31}30 (18.6\%) & \cellcolor{rqgreen!37}84 (52.2\%) & \cellcolor{rqgreen!28}29 (18.0\%) & \$0.059 & \$0.328 \\
    \cmidrule(lr){2-14}
     & \multirow{3}{*}{MopenHands~\cite{githubGitHubMultiswebenchMopenhands}} & DeepSeek-V3.2 & \makebox[1.15em][r]{1}--\makebox[1.75em][l]{100}~(77) & \cellcolor{rqgreen!60}33 (31.4\%) & \cellcolor{rqgreen!72}21 (37.5\%) & \cellcolor{rqgreen!60}34 (29.8\%) & \cellcolor{rqgreen!66}20 (42.6\%) & \cellcolor{rqgreen!69}68 (42.2\%) & \cellcolor{rqgreen!72}55 (34.2\%) & \cellcolor{rqgreen!68}118 (73.3\%) & \cellcolor{rqgreen!66}54 (33.5\%) & \$0.141 & \$0.421 \\
     &  & Qwen3.5-Plus & \makebox[1.15em][r]{14}--\makebox[1.75em][l]{100}~(94) & \cellcolor{rqgreen!76}40 (38.1\%) & \cellcolor{rqgreen!80}\textbf{23 (41.1\%)} & \cellcolor{rqgreen!80}\textbf{44 (38.6\%)} & \cellcolor{rqgreen!62}19 (40.4\%) & \cellcolor{rqgreen!72}71 (44.1\%) & \cellcolor{rqgreen!75}57 (35.4\%) & \cellcolor{rqgreen!65}115 (71.4\%) & \cellcolor{rqgreen!80}\textbf{63 (39.1\%)} & \$0.232 & \$0.594 \\
     &  & GPT-5.6 Luna & \makebox[1.15em][r]{4}--\makebox[1.75em][l]{100}~(28) & \cellcolor{rqgreen!51}29 (27.6\%) & \cellcolor{rqgreen!67}20 (35.7\%) & \cellcolor{rqgreen!54}31 (27.2\%) & \cellcolor{rqgreen!58}18 (38.3\%) & \cellcolor{rqgreen!59}60 (37.3\%) & \cellcolor{rqgreen!67}52 (32.3\%) & \cellcolor{rqgreen!70}120 (74.5\%) & \cellcolor{rqgreen!59}49 (30.4\%) & \$0.023 & \$0.074 \\
    \cmidrule(lr){2-14}
     & \multirow{3}{*}{mini-swe-agent~\cite{githubGitHubSWEagentminisweagent}} & DeepSeek-V3.2 & \makebox[1.15em][r]{28}--\makebox[1.75em][l]{100}~(73) & \cellcolor{rqgreen!73}39 (37.1\%) & \cellcolor{rqgreen!55}17 (30.4\%) & \cellcolor{rqgreen!68}38 (33.3\%) & \cellcolor{rqgreen!58}18 (38.3\%) & \cellcolor{rqgreen!65}65 (40.4\%) & \cellcolor{rqgreen!77}58 (36.0\%) & \cellcolor{rqgreen!59}108 (67.1\%) & \cellcolor{rqgreen!69}56 (34.8\%) & \$0.099 & \$0.284 \\
     &  & Qwen3.5-Plus & \makebox[1.15em][r]{27}--\makebox[1.75em][l]{100}~(74) & \cellcolor{rqgreen!66}36 (34.3\%) & \cellcolor{rqgreen!59}18 (32.1\%) & \cellcolor{rqgreen!62}35 (30.7\%) & \cellcolor{rqgreen!62}19 (40.4\%) & \cellcolor{rqgreen!64}64 (39.8\%) & \cellcolor{rqgreen!69}53 (32.9\%) & \cellcolor{rqgreen!51}99 (61.5\%) & \cellcolor{rqgreen!66}54 (33.5\%) & \$0.114 & \$0.341 \\
     &  & GPT-5.6 Luna & \makebox[1.15em][r]{4}--\makebox[1.75em][l]{26}~(10) & \cellcolor{rqgreen!66}36 (34.3\%) & \cellcolor{rqgreen!67}20 (35.7\%) & \cellcolor{rqgreen!60}34 (29.8\%) & \cellcolor{rqgreen!76}22 (46.8\%) & \cellcolor{rqgreen!65}65 (40.4\%) & \cellcolor{rqgreen!67}52 (32.3\%) & \cellcolor{rqgreen!80}\textbf{131 (81.4\%)} & \cellcolor{rqgreen!69}56 (34.8\%) & \textbf{\$0.007} & \textbf{\$0.019} \\
    \midrule\midrule
    \multirow{6}{*}{\rotatebox{90}{Coding agents}} & \multirow{3}{*}{OpenCode~\cite{githubGitHubAnomalycoopencode}} & DeepSeek-V3.2 & \makebox[1.15em][r]{28}--\makebox[1.75em][l]{100}~(58) & \cellcolor{rqgreen!80}\textbf{42 (40.0\%)} & \cellcolor{rqgreen!67}20 (35.7\%) & \cellcolor{rqgreen!70}39 (34.2\%) & \cellcolor{rqgreen!80}\textbf{23 (48.9\%)} & \cellcolor{rqgreen!80}\textbf{77 (47.8\%)} & \cellcolor{rqgreen!80}\textbf{60 (37.3\%)} & \cellcolor{rqgreen!71}121 (75.2\%) & \cellcolor{rqgreen!78}62 (38.5\%) & \$0.165 & \$0.429 \\
     &  & Qwen3.5-Plus & \makebox[1.15em][r]{11}--\makebox[1.75em][l]{100}~(61) & \cellcolor{rqgreen!71}38 (36.2\%) & \cellcolor{rqgreen!55}17 (30.4\%) & \cellcolor{rqgreen!64}36 (31.6\%) & \cellcolor{rqgreen!62}19 (40.4\%) & \cellcolor{rqgreen!62}63 (39.1\%) & \cellcolor{rqgreen!62}49 (30.4\%) & \cellcolor{rqgreen!67}117 (72.7\%) & \cellcolor{rqgreen!68}55 (34.2\%) & \$0.160 & \$0.468 \\
     &  & GPT-5.6 Luna & \makebox[1.15em][r]{10}--\makebox[1.75em][l]{100}~(26) & \cellcolor{rqgreen!49}28 (26.7\%) & \cellcolor{rqgreen!76}22 (39.3\%) & \cellcolor{rqgreen!60}34 (29.8\%) & \cellcolor{rqgreen!48}16 (34.0\%) & \cellcolor{rqgreen!56}58 (36.0\%) & \cellcolor{rqgreen!60}48 (29.8\%) & \cellcolor{rqgreen!52}100 (62.1\%) & \cellcolor{rqgreen!60}50 (31.1\%) & \$0.022 & \$0.071 \\
    \cmidrule(lr){2-14}
     & \multirow{3}{*}{Codex~\cite{githubGitHubOpenaicodex}} & DeepSeek-V3.2 & \makebox[1.15em][r]{8}--\makebox[1.75em][l]{100}~(67) & \cellcolor{rqgreen!64}35 (33.3\%) & \cellcolor{rqgreen!59}18 (32.1\%) & \cellcolor{rqgreen!64}36 (31.6\%) & \cellcolor{rqgreen!53}17 (36.2\%) & \cellcolor{rqgreen!67}67 (41.6\%) & \cellcolor{rqgreen!69}53 (32.9\%) & \cellcolor{rqgreen!57}106 (65.8\%) & \cellcolor{rqgreen!65}53 (32.9\%) & \$0.120 & \$0.366 \\
     &  & Qwen3.5-Plus & \makebox[1.15em][r]{6}--\makebox[1.75em][l]{100}~(56) & \cellcolor{rqgreen!69}37 (35.2\%) & \cellcolor{rqgreen!72}21 (37.5\%) & \cellcolor{rqgreen!70}39 (34.2\%) & \cellcolor{rqgreen!62}19 (40.4\%) & \cellcolor{rqgreen!70}69 (42.9\%) & \cellcolor{rqgreen!69}53 (32.9\%) & \cellcolor{rqgreen!65}115 (71.4\%) & \cellcolor{rqgreen!72}58 (36.0\%) & \$0.122 & \$0.340 \\
     &  & GPT-5.6 Luna & \makebox[1.15em][r]{1}--\makebox[1.75em][l]{70}~(14) & \cellcolor{rqgreen!69}37 (35.2\%) & \cellcolor{rqgreen!72}21 (37.5\%) & \cellcolor{rqgreen!68}38 (33.3\%) & \cellcolor{rqgreen!66}20 (42.6\%) & \cellcolor{rqgreen!66}66 (41.0\%) & \cellcolor{rqgreen!65}51 (31.7\%) & \cellcolor{rqgreen!67}117 (72.7\%) & \cellcolor{rqgreen!72}58 (36.0\%) & \$0.042 & \$0.118 \\
    \bottomrule
  \end{tabular}}
\end{table}

%% file: tables/rq1_group3.tex
\begin{table}[t]
  \centering
  \footnotesize
  \caption{Repair performance under mini-swe-agent. Models marked $\dagger$ are newly evaluated and postdate the leakage boundary; steps are min--max (mean).}
  \label{tab:rq1group3}
  \setlength{\tabcolsep}{3.2pt}
  \resizebox{\linewidth}{!}{%
  \begin{tabular}{@{}l r r r r r r r@{}}
    \toprule
    Model & DeepSeek-V4.1-Flash$^\dagger$ & GLM-5.3-Flash$^\dagger$ & Qwen3.8-Flash$^\dagger$ & GPT-6 Luna$^\dagger$ & DeepSeek-V3.2 & GPT-5.6 Luna & Qwen3.5-Plus \\
    \midrule
    \#Resolved (\%Resolved) & 65 (40.4\%) & 65 (40.4\%) & 58 (36.0\%) & 58 (36.0\%) & 56 (34.8\%) & 56 (34.8\%) & 54 (33.5\%) \\
    \quad Python / Java & 37.7\% / 46.8\% & 36.0\% / 51.1\% & 34.2\% / 40.4\% & 31.6\% / 46.8\% & 33.3\% / 38.3\% & 29.8\% / 46.8\% & 30.7\% / 40.4\% \\
    \midrule
    Avg.\ \$ / \$/Solved & \$0.030 / \$0.073 & \$0.023 / \$0.056 & \$0.030 / \$0.082 & \$0.004 / \$0.012 & \$0.099 / \$0.284 & \$0.007 / \$0.019 & \$0.114 / \$0.341 \\
    Steps (min--max, mean) & 6--100 (49) & 5--100 (29) & 6--100 (55) & 6--85 (17) & 28--100 (73) & 4--26 (10) & 27--100 (74) \\
    \bottomrule
  \end{tabular}}
\end{table}

%% file: tables/rq2.tex
\definecolor{rqgreen}{HTML}{27AE60}
\begin{table}[t]
  \centering
  \footnotesize
  \caption{Repair performance by defect characteristic across 18 configurations. \emph{Py}/\emph{Java} and $\le$02-16/$>$02-16 denote language and issue-date splits; \emph{Others} groups values with fewer than five instances.}
  \label{tab:rq2}
  \setlength{\tabcolsep}{1.4pt}
  \resizebox{\linewidth}{!}{%
  \begin{tabular}{@{}l r r r r r >{\centering\arraybackslash}m{3.2em} >{\centering\arraybackslash}m{3.2em} >{\centering\arraybackslash}m{3.2em} >{\centering\arraybackslash}m{3.2em} >{\centering\arraybackslash}m{3.2em} >{\centering\arraybackslash}m{3.2em} >{\centering\arraybackslash}m{3.2em} >{\centering\arraybackslash}m{3.2em} >{\centering\arraybackslash}m{3.2em} >{\centering\arraybackslash}m{3.2em} >{\centering\arraybackslash}m{3.2em} >{\centering\arraybackslash}m{3.2em} >{\centering\arraybackslash}m{3.2em} >{\centering\arraybackslash}m{3.2em} >{\centering\arraybackslash}m{3.2em} >{\centering\arraybackslash}m{3.2em} >{\centering\arraybackslash}m{3.2em} >{\centering\arraybackslash}m{3.2em} r r r l@{}}
    \toprule
    \multirow{2}{*}[-2.3pt]{\textbf{Characteristic}} & \multirow{2}{*}[-2.3pt]{\textbf{\#Inst}} & \multirow{2}{*}[-2.3pt]{\textbf{Py}} & \multirow{2}{*}[-2.3pt]{\textbf{Java}} & \multirow{2}{*}[-2.3pt]{\textbf{$\le$02-16}} & \multirow{2}{*}[-2.3pt]{\textbf{$>$02-16}} & \multicolumn{3}{c}{\textbf{MSWE-agent}} & \multicolumn{3}{c}{\textbf{Magentless}} & \multicolumn{3}{c}{\textbf{MopenHands}} & \multicolumn{3}{c}{\textbf{mini-swe-agent}} & \multicolumn{3}{c}{\textbf{OpenCode}} & \multicolumn{3}{c}{\textbf{Codex}} & \multirow{2}{*}[-2.3pt]{\textbf{min}} & \multirow{2}{*}[-2.3pt]{\textbf{max}} & \multirow{2}{*}[-2.3pt]{\textbf{$\ge$1}} & \multirow{2}{*}[-2.3pt]{\textbf{Avg}} \\
    \cmidrule(lr){7-9} \cmidrule(lr){10-12} \cmidrule(lr){13-15} \cmidrule(lr){16-18} \cmidrule(lr){19-21} \cmidrule(lr){22-24}
     &  &  &  &  &  & DS & Qwen & GPT & DS & Qwen & GPT & DS & Qwen & GPT & DS & Qwen & GPT & DS & Qwen & GPT & DS & Qwen & GPT &  &  &  &  \\
    \midrule
    \multicolumn{28}{@{}l}{\cellcolor{black!10}\textit{Sensitive information type}}
    \\
    credential\_secret & 70 & 30.6\% & 37.8\% & 28.5\% & 41.5\% & 30.0\% & 27.1\% & 30.0\% & 20.0\% & 14.3\% & 15.7\% & 38.6\% & \textbf{44.3\%} & 28.6\% & 41.4\% & 35.7\% & 35.7\% & 41.4\% & 38.6\% & 30.0\% & 35.7\% & 41.4\% & 41.4\% & 14.3\% & 44.3\% & 64.3\% & \databar{32.8}{100}{rqgreen!80}{32.8\%} \\
    \rowcolor{black!4}private\_record & 45 & 22.6\% & 38.5\% & 29.5\% & 23.6\% & 26.7\% & 31.1\% & 28.9\% & 13.3\% & 11.1\% & 15.6\% & \textbf{33.3\%} & 28.9\% & \textbf{33.3\%} & 31.1\% & 31.1\% & \textbf{33.3\%} & \textbf{33.3\%} & 31.1\% & 31.1\% & 31.1\% & 28.9\% & 28.9\% & 11.1\% & 33.3\% & 55.6\% & \databar{27.9}{100}{rqgreen!80}{27.9\%} \\
    business\_record & 22 & 33.3\% & 35.7\% & 35.4\% & 30.6\% & 27.3\% & 31.8\% & 40.9\% & 13.6\% & 4.5\% & 22.7\% & 27.3\% & \textbf{54.5\%} & 36.4\% & 40.9\% & 36.4\% & 40.9\% & \textbf{54.5\%} & 40.9\% & 27.3\% & 36.4\% & 40.9\% & 36.4\% & 4.5\% & 54.5\% & 68.2\% & \databar{34.1}{100}{rqgreen!80}{34.1\%} \\
    \rowcolor{black!4}file\_content & 13 & 23.1\% & 0.0\% & -- & 21.4\% & 7.7\% & 15.4\% & 23.1\% & 7.7\% & 0.0\% & 23.1\% & 15.4\% & 30.8\% & 23.1\% & 23.1\% & 30.8\% & 23.1\% & 23.1\% & 23.1\% & \textbf{38.5\%} & 7.7\% & 30.8\% & \textbf{38.5\%} & 0.0\% & 38.5\% & 38.5\% & \databar{21.4}{100}{rqgreen!80}{21.4\%} \\
    file\_path & 7 & 6.9\% & 38.9\% & 21.3\% & 16.7\% & 14.3\% & 28.6\% & \textbf{42.9\%} & 0.0\% & 0.0\% & 28.6\% & \textbf{42.9\%} & 14.3\% & 14.3\% & 0.0\% & 14.3\% & 28.6\% & 28.6\% & 14.3\% & 28.6\% & \textbf{42.9\%} & 14.3\% & 14.3\% & 0.0\% & 42.9\% & 57.1\% & \databar{20.6}{100}{rqgreen!80}{20.6\%} \\
    \rowcolor{black!4}Others & 4 & 37.5\% & -- & 31.5\% & 55.6\% & 25.0\% & \textbf{50.0\%} & \textbf{50.0\%} & 25.0\% & 0.0\% & 25.0\% & 25.0\% & \textbf{50.0\%} & \textbf{50.0\%} & 25.0\% & \textbf{50.0\%} & \textbf{50.0\%} & 25.0\% & 25.0\% & \textbf{50.0\%} & \textbf{50.0\%} & \textbf{50.0\%} & \textbf{50.0\%} & 0.0\% & 50.0\% & 50.0\% & \databar{37.5}{100}{rqgreen!80}{37.5\%} \\
    \midrule
    \multicolumn{28}{@{}l}{\cellcolor{black!10}\textit{Unauthorized actor}}
    \\
    ui\_viewer & 33 & 30.2\% & 40.7\% & 34.0\% & 24.1\% & 27.3\% & 30.3\% & 30.3\% & 21.2\% & 12.1\% & 21.2\% & 36.4\% & 45.5\% & 27.3\% & 42.4\% & 39.4\% & 33.3\% & \textbf{48.5\%} & 36.4\% & 30.3\% & 39.4\% & 33.3\% & 24.2\% & 12.1\% & 48.5\% & 66.7\% & \databar{32.2}{100}{rqgreen!80}{32.2\%} \\
    \rowcolor{black!4}other\_authenticated\_user & 29 & 19.4\% & 27.8\% & 26.8\% & 16.7\% & 17.2\% & 13.8\% & 17.2\% & 10.3\% & 6.9\% & 13.8\% & 17.2\% & 31.0\% & 27.6\% & 24.1\% & 24.1\% & 31.0\% & \textbf{37.9\%} & 27.6\% & 27.6\% & 24.1\% & 27.6\% & 27.6\% & 6.9\% & 37.9\% & 55.2\% & \databar{22.6}{100}{rqgreen!80}{22.6\%} \\
    log\_reader & 29 & 36.0\% & 63.9\% & 36.9\% & 58.6\% & 44.8\% & 37.9\% & 34.5\% & 31.0\% & 17.2\% & 24.1\% & 48.3\% & 55.2\% & 44.8\% & 51.7\% & 48.3\% & 48.3\% & 51.7\% & 44.8\% & 44.8\% & 44.8\% & 55.2\% & \textbf{58.6\%} & 17.2\% & 58.6\% & 75.9\% & \databar{43.7}{100}{rqgreen!80}{43.7\%} \\
    \rowcolor{black!4}adjacent\_client & 16 & 14.8\% & 31.0\% & 27.2\% & 15.1\% & 18.8\% & 25.0\% & 31.2\% & 6.2\% & 6.2\% & 12.5\% & 31.2\% & 18.8\% & 18.8\% & 25.0\% & 18.8\% & 25.0\% & \textbf{37.5\%} & 18.8\% & 25.0\% & 31.2\% & 25.0\% & 18.8\% & 6.2\% & 37.5\% & 37.5\% & \databar{21.9}{100}{rqgreen!80}{21.9\%} \\
    cross\_tenant\_actor & 12 & 14.6\% & 51.4\% & 13.3\% & 36.5\% & 25.0\% & 33.3\% & 33.3\% & 8.3\% & 8.3\% & 8.3\% & \textbf{50.0\%} & 33.3\% & 41.7\% & 16.7\% & 33.3\% & 33.3\% & 25.0\% & 25.0\% & 25.0\% & 25.0\% & 33.3\% & 25.0\% & 8.3\% & 50.0\% & 50.0\% & \databar{26.9}{100}{rqgreen!80}{26.9\%} \\
    \rowcolor{black!4}infra\_artifact\_reader & 11 & 28.4\% & 0.0\% & 11.1\% & 44.4\% & 9.1\% & 27.3\% & 18.2\% & 18.2\% & 9.1\% & 9.1\% & 27.3\% & 27.3\% & 27.3\% & 27.3\% & 27.3\% & 27.3\% & 27.3\% & \textbf{36.4\%} & 18.2\% & 18.2\% & \textbf{36.4\%} & 27.3\% & 9.1\% & 36.4\% & 36.4\% & \databar{23.2}{100}{rqgreen!80}{23.2\%} \\
    chat\_recipient & 9 & 35.8\% & -- & 35.7\% & 36.1\% & 22.2\% & 44.4\% & \textbf{66.7\%} & 0.0\% & 0.0\% & 33.3\% & 33.3\% & 44.4\% & 22.2\% & 33.3\% & 22.2\% & 44.4\% & 33.3\% & 44.4\% & 55.6\% & 55.6\% & 33.3\% & 55.6\% & 0.0\% & 66.7\% & 77.8\% & \databar{35.8}{100}{rqgreen!80}{35.8\%} \\
    \rowcolor{black!4}workspace\_restricted\_user & 7 & 37.3\% & -- & -- & 37.3\% & 14.3\% & 42.9\% & \textbf{57.1\%} & 14.3\% & 0.0\% & 28.6\% & 14.3\% & \textbf{57.1\%} & 42.9\% & 42.9\% & \textbf{57.1\%} & 42.9\% & 28.6\% & 28.6\% & \textbf{57.1\%} & 28.6\% & \textbf{57.1\%} & \textbf{57.1\%} & 0.0\% & 57.1\% & 57.1\% & \databar{37.3}{100}{rqgreen!80}{37.3\%} \\
    redirect\_target & 7 & 23.6\% & 37.0\% & 35.6\% & 13.9\% & 28.6\% & 28.6\% & 28.6\% & 14.3\% & 28.6\% & 28.6\% & 42.9\% & 28.6\% & 28.6\% & 42.9\% & 28.6\% & 14.3\% & 28.6\% & 42.9\% & 0.0\% & 28.6\% & 28.6\% & \textbf{57.1\%} & 0.0\% & 57.1\% & 57.1\% & \databar{29.4}{100}{rqgreen!80}{29.4\%} \\
    \rowcolor{black!4}sql\_user & 7 & 27.8\% & 24.1\% & 24.6\% & -- & \textbf{42.9\%} & 14.3\% & \textbf{42.9\%} & 0.0\% & 0.0\% & 0.0\% & 28.6\% & \textbf{42.9\%} & 14.3\% & 28.6\% & 28.6\% & \textbf{42.9\%} & 14.3\% & \textbf{42.9\%} & 14.3\% & 14.3\% & 28.6\% & \textbf{42.9\%} & 0.0\% & 42.9\% & 71.4\% & \databar{24.6}{100}{rqgreen!80}{24.6\%} \\
    Others & 1 & 0.0\% & -- & 0.0\% & -- & 0.0\% & 0.0\% & 0.0\% & 0.0\% & 0.0\% & 0.0\% & 0.0\% & 0.0\% & 0.0\% & 0.0\% & 0.0\% & 0.0\% & 0.0\% & 0.0\% & 0.0\% & 0.0\% & 0.0\% & 0.0\% & 0.0\% & 0.0\% & 0.0\% & \databar{0.0}{100}{rqgreen!80}{0.0\%} \\
    \midrule
    \multicolumn{28}{@{}l}{\cellcolor{black!10}\textit{Exposure channel}}
    \\
    http\_response & 41 & 19.9\% & 32.5\% & 24.5\% & 27.5\% & 22.0\% & 22.0\% & 26.8\% & 9.8\% & 7.3\% & 14.6\% & 24.4\% & 31.7\% & 36.6\% & 22.0\% & 31.7\% & 31.7\% & \textbf{39.0\%} & 31.7\% & 24.4\% & 31.7\% & 31.7\% & 24.4\% & 7.3\% & 39.0\% & 53.7\% & \databar{25.7}{100}{rqgreen!80}{25.7\%} \\
    \rowcolor{black!4}ui & 29 & 28.9\% & 36.7\% & 32.7\% & 9.3\% & 27.6\% & 27.6\% & 27.6\% & 20.7\% & 13.8\% & 17.2\% & 27.6\% & 41.4\% & 20.7\% & 37.9\% & 41.4\% & 34.5\% & \textbf{44.8\%} & 34.5\% & 31.0\% & 31.0\% & 31.0\% & 34.5\% & 13.8\% & 44.8\% & 65.5\% & \databar{30.3}{100}{rqgreen!80}{30.3\%} \\
    log & 26 & 34.0\% & 63.9\% & 37.0\% & 56.9\% & 42.3\% & 38.5\% & 34.6\% & 30.8\% & 15.4\% & 26.9\% & 50.0\% & 53.8\% & 42.3\% & 53.8\% & 46.2\% & 46.2\% & 53.8\% & 42.3\% & 46.2\% & 42.3\% & 53.8\% & \textbf{57.7\%} & 15.4\% & 57.7\% & 76.9\% & \databar{43.2}{100}{rqgreen!80}{43.2\%} \\
    \rowcolor{black!4}tool\_result & 14 & 24.2\% & -- & 5.6\% & 29.3\% & 7.1\% & 21.4\% & 35.7\% & 7.1\% & 0.0\% & 21.4\% & \textbf{42.9\%} & 35.7\% & 21.4\% & 35.7\% & 21.4\% & 21.4\% & 28.6\% & 21.4\% & 35.7\% & 14.3\% & 28.6\% & 35.7\% & 0.0\% & 42.9\% & 57.1\% & \databar{24.2}{100}{rqgreen!80}{24.2\%} \\
    chat\_message & 12 & 18.5\% & -- & 29.4\% & 3.3\% & 16.7\% & 16.7\% & \textbf{33.3\%} & 0.0\% & 0.0\% & 25.0\% & 16.7\% & 25.0\% & 16.7\% & 25.0\% & 16.7\% & 16.7\% & 25.0\% & 8.3\% & 25.0\% & \textbf{33.3\%} & 16.7\% & 16.7\% & 0.0\% & 33.3\% & 41.7\% & \databar{18.5}{100}{rqgreen!80}{18.5\%} \\
    \rowcolor{black!4}deployment\_config & 12 & 28.3\% & 0.0\% & 13.2\% & 44.4\% & 8.3\% & 33.3\% & 25.0\% & 16.7\% & 8.3\% & 8.3\% & 25.0\% & 33.3\% & 25.0\% & 25.0\% & 25.0\% & 33.3\% & 25.0\% & \textbf{41.7\%} & 16.7\% & 16.7\% & 33.3\% & 25.0\% & 8.3\% & 41.7\% & 41.7\% & \databar{23.6}{100}{rqgreen!80}{23.6\%} \\
    http\_request\_header & 8 & 23.6\% & 38.9\% & 35.7\% & 0.0\% & 25.0\% & 37.5\% & 25.0\% & 12.5\% & 25.0\% & 25.0\% & \textbf{50.0\%} & 25.0\% & 25.0\% & \textbf{50.0\%} & 25.0\% & 25.0\% & 37.5\% & \textbf{50.0\%} & 12.5\% & 37.5\% & 25.0\% & \textbf{50.0\%} & 12.5\% & 50.0\% & 50.0\% & \databar{31.2}{100}{rqgreen!80}{31.2\%} \\
    \rowcolor{black!4}sql\_result & 7 & 27.8\% & 24.1\% & 24.6\% & -- & \textbf{42.9\%} & 14.3\% & \textbf{42.9\%} & 0.0\% & 0.0\% & 0.0\% & 28.6\% & \textbf{42.9\%} & 14.3\% & 28.6\% & 28.6\% & \textbf{42.9\%} & 14.3\% & \textbf{42.9\%} & 14.3\% & 14.3\% & 28.6\% & \textbf{42.9\%} & 0.0\% & 42.9\% & 71.4\% & \databar{24.6}{100}{rqgreen!80}{24.6\%} \\
    error\_message & 6 & 48.1\% & 42.6\% & 66.7\% & 24.1\% & 50.0\% & 50.0\% & 33.3\% & 16.7\% & 16.7\% & 0.0\% & 50.0\% & \textbf{66.7\%} & 50.0\% & 33.3\% & \textbf{66.7\%} & \textbf{66.7\%} & 50.0\% & 50.0\% & 50.0\% & \textbf{66.7\%} & \textbf{66.7\%} & 33.3\% & 0.0\% & 66.7\% & 66.7\% & \databar{45.4}{100}{rqgreen!80}{45.4\%} \\
    \rowcolor{black!4}Others & 6 & 46.3\% & -- & 22.2\% & 58.3\% & 33.3\% & 50.0\% & \textbf{66.7\%} & 33.3\% & 16.7\% & 33.3\% & 50.0\% & 50.0\% & 50.0\% & 50.0\% & 16.7\% & 50.0\% & 33.3\% & 33.3\% & \textbf{66.7\%} & \textbf{66.7\%} & \textbf{66.7\%} & \textbf{66.7\%} & 16.7\% & 66.7\% & 66.7\% & \databar{46.3}{100}{rqgreen!80}{46.3\%} \\
    \midrule
    \multicolumn{28}{@{}l}{\cellcolor{black!10}\textit{Repair pattern}}
    \\
    mask\_redact & 58 & 30.7\% & 50.6\% & 37.1\% & 38.9\% & 34.5\% & 31.0\% & 36.2\% & 19.0\% & 13.8\% & 20.7\% & 43.1\% & \textbf{51.7\%} & 32.8\% & 46.6\% & 41.4\% & 43.1\% & 48.3\% & 43.1\% & 37.9\% & 41.4\% & 44.8\% & 46.6\% & 13.8\% & 51.7\% & 72.4\% & \databar{37.5}{100}{rqgreen!80}{37.5\%} \\
    \rowcolor{black!4}add\_authorization\_check & 47 & 25.1\% & 30.6\% & 24.7\% & 28.5\% & 21.3\% & 25.5\% & 25.5\% & 12.8\% & 4.3\% & 14.9\% & 23.4\% & \textbf{38.3\%} & 34.0\% & 25.5\% & 34.0\% & 31.9\% & \textbf{38.3\%} & 29.8\% & 25.5\% & 29.8\% & 31.9\% & 29.8\% & 4.3\% & 38.3\% & 53.2\% & \databar{26.5}{100}{rqgreen!80}{26.5\%} \\
    remove\_field & 36 & 27.8\% & 38.9\% & 25.6\% & 42.2\% & 30.6\% & 33.3\% & 33.3\% & 19.4\% & 11.1\% & 16.7\% & 33.3\% & 30.6\% & 27.8\% & 36.1\% & 30.6\% & 36.1\% & 36.1\% & 27.8\% & 36.1\% & 36.1\% & \textbf{38.9\%} & 30.6\% & 11.1\% & 38.9\% & 55.6\% & \databar{30.2}{100}{rqgreen!80}{30.2\%} \\
    \rowcolor{black!4}defer\_resolution & 7 & 40.7\% & 0.0\% & 24.4\% & 0.0\% & 14.3\% & \textbf{28.6\%} & \textbf{28.6\%} & 14.3\% & 14.3\% & 0.0\% & 14.3\% & \textbf{28.6\%} & 14.3\% & 14.3\% & 14.3\% & \textbf{28.6\%} & 14.3\% & \textbf{28.6\%} & 14.3\% & 14.3\% & 14.3\% & 14.3\% & 0.0\% & 28.6\% & 28.6\% & \databar{17.5}{100}{rqgreen!80}{17.5\%} \\
    reject\_hardlink & 6 & 26.9\% & -- & 66.7\% & 18.9\% & 0.0\% & 33.3\% & 33.3\% & 0.0\% & 0.0\% & 33.3\% & \textbf{50.0\%} & 33.3\% & 33.3\% & 33.3\% & 33.3\% & 16.7\% & 33.3\% & 33.3\% & 33.3\% & 0.0\% & 33.3\% & \textbf{50.0\%} & 0.0\% & 50.0\% & 50.0\% & \databar{26.9}{100}{rqgreen!80}{26.9\%} \\
    \rowcolor{black!4}Others & 7 & 6.9\% & 16.7\% & 6.7\% & 22.2\% & 0.0\% & 0.0\% & \textbf{28.6\%} & 0.0\% & 14.3\% & \textbf{28.6\%} & \textbf{28.6\%} & 0.0\% & 14.3\% & 14.3\% & 0.0\% & 0.0\% & 0.0\% & \textbf{28.6\%} & 0.0\% & 14.3\% & 0.0\% & \textbf{28.6\%} & 0.0\% & 28.6\% & 57.1\% & \databar{11.1}{100}{rqgreen!80}{11.1\%} \\
    \midrule
    \multicolumn{28}{@{}l}{\cellcolor{black!10}\textit{Primary CWE}}
    \\
    CWE-201 & 51 & 35.4\% & 27.4\% & 33.8\% & 31.9\% & 29.4\% & 29.4\% & 39.2\% & 17.6\% & 11.8\% & 21.6\% & 39.2\% & 41.2\% & 27.5\% & 43.1\% & 31.4\% & 37.3\% & 45.1\% & 33.3\% & 33.3\% & \textbf{47.1\%} & 35.3\% & 35.3\% & 11.8\% & 47.1\% & 58.8\% & \databar{33.2}{100}{rqgreen!80}{33.2\%} \\
    \rowcolor{black!4}CWE-862 & 36 & 18.6\% & 29.9\% & 22.2\% & 23.6\% & 19.4\% & 25.0\% & 25.0\% & 5.6\% & 2.8\% & 11.1\% & 22.2\% & 30.6\% & \textbf{33.3\%} & 16.7\% & 25.0\% & 30.6\% & \textbf{33.3\%} & 27.8\% & 25.0\% & 25.0\% & 25.0\% & 25.0\% & 2.8\% & 33.3\% & 58.3\% & \databar{22.7}{100}{rqgreen!80}{22.7\%} \\
    CWE-532 & 26 & 34.0\% & 63.9\% & 37.0\% & 56.9\% & 42.3\% & 38.5\% & 34.6\% & 30.8\% & 15.4\% & 26.9\% & 50.0\% & 53.8\% & 42.3\% & 53.8\% & 46.2\% & 46.2\% & 53.8\% & 42.3\% & 46.2\% & 42.3\% & 53.8\% & \textbf{57.7\%} & 15.4\% & 57.7\% & 76.9\% & \databar{43.2}{100}{rqgreen!80}{43.2\%} \\
    \rowcolor{black!4}CWE-538 & 12 & 16.2\% & 0.0\% & 8.3\% & 47.2\% & 8.3\% & 16.7\% & 16.7\% & 8.3\% & 8.3\% & 0.0\% & 16.7\% & 25.0\% & 8.3\% & 16.7\% & 16.7\% & 25.0\% & 16.7\% & \textbf{41.7\%} & 8.3\% & 8.3\% & 16.7\% & 8.3\% & 0.0\% & 41.7\% & 50.0\% & \databar{14.8}{100}{rqgreen!80}{14.8\%} \\
    CWE-22 & 9 & 23.5\% & -- & -- & 23.5\% & 11.1\% & 11.1\% & 22.2\% & 11.1\% & 0.0\% & 22.2\% & 11.1\% & 33.3\% & 33.3\% & 22.2\% & 33.3\% & 33.3\% & 22.2\% & 22.2\% & \textbf{44.4\%} & 11.1\% & 33.3\% & \textbf{44.4\%} & 0.0\% & 44.4\% & 44.4\% & \databar{23.5}{100}{rqgreen!80}{23.5\%} \\
    \rowcolor{black!4}Others & 27 & 24.7\% & 41.4\% & 35.3\% & 25.0\% & 25.9\% & 33.3\% & 33.3\% & 14.8\% & 14.8\% & 18.5\% & 37.0\% & 40.7\% & 29.6\% & 37.0\% & \textbf{44.4\%} & 29.6\% & 33.3\% & 37.0\% & 25.9\% & 25.9\% & \textbf{44.4\%} & 40.7\% & 14.8\% & 44.4\% & 55.6\% & \databar{31.5}{100}{rqgreen!80}{31.5\%} \\
    \midrule
    \multicolumn{28}{@{}l}{\cellcolor{black!10}\textit{Resultant CWE}}
    \\
    CWE-201 & 91 & 28.0\% & 28.5\% & 30.2\% & 24.3\% & 23.1\% & 25.3\% & 33.0\% & 13.2\% & 8.8\% & 19.8\% & 28.6\% & 35.2\% & 27.5\% & 33.0\% & 28.6\% & 33.0\% & \textbf{39.6\%} & 30.8\% & 29.7\% & 35.2\% & 29.7\% & 33.0\% & 8.8\% & 39.6\% & 59.3\% & \databar{28.1}{100}{rqgreen!80}{28.1\%} \\
    \rowcolor{black!4}CWE-532 & 26 & 34.0\% & 63.9\% & 37.0\% & 56.9\% & 42.3\% & 38.5\% & 34.6\% & 30.8\% & 15.4\% & 26.9\% & 50.0\% & 53.8\% & 42.3\% & 53.8\% & 46.2\% & 46.2\% & 53.8\% & 42.3\% & 46.2\% & 42.3\% & 53.8\% & \textbf{57.7\%} & 15.4\% & 57.7\% & 76.9\% & \databar{43.2}{100}{rqgreen!80}{43.2\%} \\
    CWE-538 & 19 & 21.2\% & 13.0\% & 15.4\% & 29.6\% & 10.5\% & 26.3\% & 21.1\% & 10.5\% & 5.3\% & 5.3\% & 26.3\% & \textbf{36.8\%} & 15.8\% & 21.1\% & 31.6\% & 21.1\% & 21.1\% & \textbf{36.8\%} & 10.5\% & 10.5\% & 31.6\% & 15.8\% & 5.3\% & 36.8\% & 47.4\% & \databar{19.9}{100}{rqgreen!80}{19.9\%} \\
    \rowcolor{black!4}CWE-359 & 8 & 25.9\% & 41.1\% & 19.4\% & 51.4\% & 37.5\% & 37.5\% & 37.5\% & 12.5\% & 12.5\% & 12.5\% & 37.5\% & \textbf{50.0\%} & \textbf{50.0\%} & 37.5\% & 37.5\% & \textbf{50.0\%} & 37.5\% & 37.5\% & 37.5\% & 25.0\% & \textbf{50.0\%} & 37.5\% & 12.5\% & 50.0\% & 50.0\% & \databar{35.4}{100}{rqgreen!80}{35.4\%} \\
    Others & 17 & 22.8\% & 42.9\% & 33.8\% & 25.9\% & 29.4\% & 29.4\% & 29.4\% & 11.8\% & 11.8\% & 11.8\% & \textbf{41.2\%} & 35.3\% & 35.3\% & 29.4\% & \textbf{41.2\%} & 35.3\% & 29.4\% & 35.3\% & 35.3\% & 35.3\% & \textbf{41.2\%} & \textbf{41.2\%} & 11.8\% & 41.2\% & 52.9\% & \databar{31.0}{100}{rqgreen!80}{31.0\%} \\
    \midrule
    \multicolumn{28}{@{}l}{\cellcolor{black!10}\textit{Primary CWE scope}}
    \\
    In CWE-200 & 104 & 30.3\% & 39.8\% & 31.2\% & 37.9\% & 30.8\% & 29.8\% & 33.7\% & 19.2\% & 12.5\% & 18.3\% & 39.4\% & 41.3\% & 28.8\% & 41.3\% & 34.6\% & 37.5\% & \textbf{42.3\%} & 37.5\% & 32.7\% & 39.4\% & 38.5\% & 37.5\% & 12.5\% & 42.3\% & 60.6\% & \databar{33.1}{100}{rqgreen!80}{33.1\%} \\
    \rowcolor{black!4}Outside CWE-200 & 57 & 22.2\% & 32.0\% & 25.4\% & 24.9\% & 17.5\% & 26.3\% & 28.1\% & 8.8\% & 5.3\% & 17.5\% & 22.8\% & \textbf{35.1\%} & 33.3\% & 22.8\% & 31.6\% & 29.8\% & 31.6\% & 28.1\% & 28.1\% & 21.1\% & 31.6\% & 33.3\% & 5.3\% & 35.1\% & 57.9\% & \databar{25.1}{100}{rqgreen!80}{25.1\%} \\
    \midrule
    \multicolumn{28}{@{}l}{\cellcolor{black!10}\textit{Application domain}}
    \\
    AI \& Machine Learning & 58 & 27.4\% & 58.3\% & 26.2\% & 29.7\% & 17.2\% & 24.1\% & 29.3\% & 17.2\% & 6.9\% & 22.4\% & 34.5\% & \textbf{39.7\%} & 27.6\% & 31.0\% & 27.6\% & 29.3\% & 34.5\% & 32.8\% & 34.5\% & 32.8\% & 34.5\% & 36.2\% & 6.9\% & 39.7\% & 55.2\% & \databar{28.4}{100}{rqgreen!80}{28.4\%} \\
    \rowcolor{black!4}Data Infrastructure & 33 & 26.9\% & 28.5\% & 26.7\% & 31.5\% & 21.2\% & 27.3\% & 33.3\% & 12.1\% & 6.1\% & 9.1\% & 36.4\% & 36.4\% & 24.2\% & 30.3\% & 33.3\% & 36.4\% & \textbf{39.4\%} & 33.3\% & 24.2\% & 21.2\% & 33.3\% & \textbf{39.4\%} & 6.1\% & 39.4\% & 60.6\% & \databar{27.6}{100}{rqgreen!80}{27.6\%} \\
    Security \& Identity & 19 & 0.0\% & 29.2\% & 20.7\% & 29.9\% & 26.3\% & 26.3\% & \textbf{31.6\%} & 5.3\% & 15.8\% & 10.5\% & \textbf{31.6\%} & \textbf{31.6\%} & 26.3\% & 21.1\% & 26.3\% & \textbf{31.6\%} & 26.3\% & 21.1\% & 26.3\% & 26.3\% & \textbf{31.6\%} & 26.3\% & 5.3\% & 31.6\% & 47.4\% & \databar{24.6}{100}{rqgreen!80}{24.6\%} \\
    \rowcolor{black!4}Others & 19 & 28.9\% & -- & 27.5\% & 41.7\% & 36.8\% & 31.6\% & 26.3\% & 15.8\% & 5.3\% & 10.5\% & 21.1\% & \textbf{42.1\%} & 21.1\% & \textbf{42.1\%} & 36.8\% & 26.3\% & 36.8\% & 26.3\% & 31.6\% & 36.8\% & 36.8\% & 36.8\% & 5.3\% & 42.1\% & 63.2\% & \databar{28.9}{100}{rqgreen!80}{28.9\%} \\
    Web \& API & 15 & 50.6\% & 53.7\% & 50.4\% & 72.2\% & 60.0\% & 53.3\% & 46.7\% & 26.7\% & 20.0\% & 33.3\% & 40.0\% & 60.0\% & \textbf{73.3\%} & 60.0\% & 66.7\% & 60.0\% & \textbf{73.3\%} & 46.7\% & 46.7\% & 66.7\% & 60.0\% & 40.0\% & 20.0\% & 73.3\% & 80.0\% & \databar{51.9}{100}{rqgreen!80}{51.9\%} \\
    \rowcolor{black!4}DevOps \& Cloud & 12 & 12.5\% & 37.5\% & 22.7\% & 0.0\% & 16.7\% & 8.3\% & 16.7\% & 16.7\% & 8.3\% & 8.3\% & 16.7\% & \textbf{33.3\%} & 25.0\% & \textbf{33.3\%} & 25.0\% & \textbf{33.3\%} & 25.0\% & \textbf{33.3\%} & 8.3\% & 25.0\% & 25.0\% & 16.7\% & 8.3\% & 33.3\% & 50.0\% & \databar{20.8}{100}{rqgreen!80}{20.8\%} \\
    Developer Tools & 5 & 11.1\% & 63.9\% & 44.4\% & 88.9\% & 40.0\% & 60.0\% & 60.0\% & 20.0\% & 40.0\% & 60.0\% & 80.0\% & 20.0\% & 40.0\% & 60.0\% & 40.0\% & 60.0\% & 60.0\% & \textbf{100.0\%} & 60.0\% & 40.0\% & 40.0\% & 80.0\% & 20.0\% & 100.0\% & 100.0\% & \databar{53.3}{100}{rqgreen!80}{53.3\%} \\
    \bottomrule
  \end{tabular}}
\end{table}

%% file: tables/rq3_failure_causes.tex
\definecolor{scopegreen}{HTML}{27AE60}
\definecolor{failred}{HTML}{C0392B}
\definecolor{scopeorange}{HTML}{E67E22}
\definecolor{scopeblue}{HTML}{7F8C8D}
\begin{table}[t]
  \centering
  \footnotesize
  \caption{Failure-cause distributions across the 18 configurations. Values are percentages of analyzed failures within each column. Scope indicates the generality of each cause: generic to issue resolution (\protect\tagchip{scopeblue}{any\ fix}\hspace{-0.18em}), generic to security repair (\protect\tagchip{scopeorange}{sec.\ fix}\hspace{-0.18em}), or specific to \infoex repair (\protect\tagchip{scopegreen}{\infoex}\hspace{-0.18em}).}
  \label{tab:rq3causes}
  \setlength{\tabcolsep}{0.45pt}
  \setlength{\databarw}{4.6em}
  \resizebox{\linewidth}{!}{%
  \begin{tabular}{@{}l l @{\hspace{5pt}} l r r r r >{\raggedleft\arraybackslash}m{2.75em} >{\raggedleft\arraybackslash}m{2.75em} >{\raggedleft\arraybackslash}m{2.75em} >{\raggedleft\arraybackslash}m{2.75em} >{\raggedleft\arraybackslash}m{2.75em} >{\raggedleft\arraybackslash}m{2.75em} >{\raggedleft\arraybackslash}m{2.75em} >{\raggedleft\arraybackslash}m{2.75em} >{\raggedleft\arraybackslash}m{2.75em} >{\raggedleft\arraybackslash}m{2.75em} >{\raggedleft\arraybackslash}m{2.75em} >{\raggedleft\arraybackslash}m{2.75em} >{\raggedleft\arraybackslash}m{2.75em} >{\raggedleft\arraybackslash}m{2.75em} >{\raggedleft\arraybackslash}m{2.75em} >{\raggedleft\arraybackslash}m{2.75em} >{\raggedleft\arraybackslash}m{2.75em} >{\raggedleft\arraybackslash}m{2.75em} @{\hspace{5pt}} r@{}}
    \toprule
    \multirow{2}{*}[-1.7pt]{\textbf{Cause}} & \multirow{2}{*}[-1.7pt]{\textbf{Description}} & \multirow{2}{*}[-1.7pt]{\textbf{Scope}} & \multirow{2}{*}[-1.7pt]{\textbf{Py}} & \multirow{2}{*}[-1.7pt]{\textbf{Java}} & \multirow{2}{*}[-1.7pt]{\textbf{$\le$02-16}} & \multirow{2}{*}[-1.7pt]{\textbf{$>$02-16}} & \multicolumn{3}{c}{\textbf{MSWE-agent}} & \multicolumn{3}{c}{\textbf{Magentless}} & \multicolumn{3}{c}{\textbf{MopenHands}} & \multicolumn{3}{c}{\textbf{mini-swe-agent}} & \multicolumn{3}{c}{\textbf{OpenCode}} & \multicolumn{3}{c}{\textbf{Codex}} & \multirow{2}{*}[-1.7pt]{\textbf{All}} \\
    \cmidrule(lr){8-10} \cmidrule(lr){11-13} \cmidrule(lr){14-16} \cmidrule(lr){17-19} \cmidrule(lr){20-22} \cmidrule(lr){23-25}
     &  &  &  &  &  &  & DS & Qwen & GPT & DS & Qwen & GPT & DS & Qwen & GPT & DS & Qwen & GPT & DS & Qwen & GPT & DS & Qwen & GPT &  \\
    \midrule
    \textbf{C1. no\_patch} & the method produced no source change & \tagchip{scopeblue}{any\ fix} & 26.7\% & 15.2\% & 24.1\% & 21.9\% & 31.9\% & 19.5\% & 21.2\% & 22.5\% & 59.2\% & 26.4\% & 11.1\% & 15.1\% & 8.3\% & 34.7\% & 40.5\% & 3.9\% & 4.2\% & 22.0\% & 25.0\% & 20.8\% & 12.3\% & 18.7\% & \databar{23.4}{100}{failred}{23.4\%} \\
    \rowcolor{black!4}\textbf{C2. patch\_invalid} & the patch does not apply or build & \tagchip{scopeblue}{any\ fix} & 7.5\% & 12.3\% & 10.1\% & 6.3\% & 22.0\% & 21.8\% & 13.8\% & 9.9\% & 12.5\% & 12.3\% & 4.2\% & 4.1\% & 7.1\% & 0.0\% & 1.3\% & 2.6\% & 5.6\% & 2.4\% & 7.5\% & 10.4\% & 13.7\% & 1.3\% & \databar{8.9}{100}{failred}{8.9\%} \\
    \textbf{C3. exposure\_not\_removed} & the sensitive value is still exposed after the patch &  & 61.7\% & 63.0\% & 59.6\% & 67.3\% & 42.9\% & 55.2\% & 61.2\% & 64.0\% & 27.5\% & 58.5\% & 72.2\% & 72.6\% & 77.4\% & 61.3\% & 48.1\% & 89.6\% & 81.9\% & 68.3\% & 58.8\% & 59.7\% & 68.5\% & 76.0\% & \databar{62.1}{100}{failred}{62.1\%} \\
    \rowcolor{black!4}C3.1. localization\_error & the patch edits code outside the leak's flow & \tagchip{scopeblue}{any\ fix} & 14.2\% & 11.1\% & 15.1\% & 9.6\% & 7.7\% & 19.5\% & 12.5\% & 18.9\% & 12.5\% & 17.9\% & 27.8\% & 17.8\% & 16.7\% & 6.7\% & 5.1\% & 11.7\% & 18.1\% & 17.1\% & 6.2\% & 10.4\% & 2.7\% & 8.0\% & \databar{13.3}{100}{failred}{13.3\%} \\
    C3.2. incomplete\_sanitization & redaction misses nested fields or is reversible & \tagchip{scopegreen}{\infoex} & 11.4\% & 13.4\% & 12.7\% & 10.4\% & 6.6\% & 8.0\% & 11.2\% & 12.6\% & 3.3\% & 13.2\% & 4.2\% & 13.7\% & 15.5\% & 13.3\% & 12.7\% & 20.8\% & 19.4\% & 15.9\% & 8.8\% & 11.7\% & 15.1\% & 14.7\% & \databar{12.0}{100}{failred}{12.0\%} \\
    \rowcolor{black!4}C3.3. another\_sink\_uncovered & the same value is still exposed through another output & \tagchip{scopegreen}{\infoex} & 5.2\% & 3.6\% & 3.9\% & 6.5\% & 0.0\% & 3.4\% & 7.5\% & 3.6\% & 1.7\% & 9.4\% & 5.6\% & 15.1\% & 4.8\% & 1.3\% & 0.0\% & 5.2\% & 4.2\% & 6.1\% & 7.5\% & 1.3\% & 4.1\% & 6.7\% & \databar{4.8}{100}{failred}{4.8\%} \\
    C3.4. secret\_in\_logs\_or\_errors & only log level or one error branch changed & \tagchip{scopegreen}{\infoex} & 6.1\% & 0.5\% & 5.3\% & 2.9\% & 5.5\% & 0.0\% & 3.8\% & 4.5\% & 0.8\% & 0.9\% & 5.6\% & 5.5\% & 7.1\% & 6.7\% & 6.3\% & 5.2\% & 2.8\% & 1.2\% & 10.0\% & 6.5\% & 5.5\% & 6.7\% & \databar{4.5}{100}{failred}{4.5\%} \\
    \rowcolor{black!4}C3.5. wrong\_value\_or\_fallback & a stale, default or wrong value reaches the consumer & \tagchip{scopegreen}{\infoex} & 3.4\% & 4.8\% & 5.2\% & 0.8\% & 4.4\% & 3.4\% & 6.2\% & 1.8\% & 0.0\% & 1.9\% & 2.8\% & 1.4\% & 4.8\% & 2.7\% & 3.8\% & 9.1\% & 4.2\% & 6.1\% & 5.0\% & 2.6\% & 4.1\% & 6.7\% & \databar{3.8}{100}{failred}{3.8\%} \\
    C3.6. unsafe\_channel\_routing & guarded content goes to the wrong recipient & \tagchip{scopegreen}{\infoex} & 1.5\% & 0.0\% & 0.0\% & 3.3\% & 0.0\% & 0.0\% & 0.0\% & 1.8\% & 0.0\% & 0.9\% & 1.4\% & 0.0\% & 2.4\% & 2.7\% & 2.5\% & 1.3\% & 1.4\% & 1.2\% & 1.2\% & 1.3\% & 1.4\% & 0.0\% & \databar{1.1}{100}{failred}{1.1\%} \\
    \rowcolor{black!4}C3.7. ineffective\_authorization & permission rule missing or wrong identity & \tagchip{scopegreen}{\infoex} & 6.4\% & 16.8\% & 9.7\% & 9.0\% & 5.5\% & 9.2\% & 8.8\% & 7.2\% & 5.0\% & 6.6\% & 13.9\% & 4.1\% & 10.7\% & 10.7\% & 7.6\% & 16.9\% & 12.5\% & 9.8\% & 7.5\% & 10.4\% & 13.7\% & 16.0\% & \databar{9.4}{100}{failred}{9.4\%} \\
    C3.8. stale\_shared\_state & cached or shared state keeps revoked access & \tagchip{scopeorange}{sec.\ fix} & 4.2\% & 3.9\% & 5.2\% & 1.8\% & 4.4\% & 1.1\% & 3.8\% & 5.4\% & 1.7\% & 3.8\% & 5.6\% & 5.5\% & 3.6\% & 6.7\% & 0.0\% & 5.2\% & 5.6\% & 4.9\% & 7.5\% & 2.6\% & 4.1\% & 4.0\% & \databar{4.1}{100}{failred}{4.1\%} \\
    \rowcolor{black!4}C3.9. path\_confinement & path checks miss links or part of the operations & \tagchip{scopeorange}{sec.\ fix} & 6.3\% & 0.7\% & 0.3\% & 13.9\% & 5.5\% & 4.6\% & 3.8\% & 5.4\% & 0.0\% & 2.8\% & 2.8\% & 4.1\% & 8.3\% & 4.0\% & 6.3\% & 6.5\% & 5.6\% & 2.4\% & 2.5\% & 10.4\% & 6.8\% & 5.3\% & \databar{4.7}{100}{failred}{4.7\%} \\
    C3.10. req\_validation & redirect, host or parameter checks are bypassable & \tagchip{scopeorange}{sec.\ fix} & 3.0\% & 8.2\% & 2.3\% & 9.0\% & 3.3\% & 5.7\% & 3.8\% & 2.7\% & 2.5\% & 0.9\% & 2.8\% & 5.5\% & 3.6\% & 6.7\% & 3.8\% & 7.8\% & 8.3\% & 3.7\% & 2.5\% & 2.6\% & 11.0\% & 8.0\% & \databar{4.5}{100}{failred}{4.5\%} \\
    \rowcolor{black!4}\textbf{C4. regression\_over\_fix} & exposure fixed, other behaviour broken &  & 3.6\% & 9.5\% & 5.8\% & 4.5\% & 3.3\% & 3.4\% & 3.8\% & 2.7\% & 0.8\% & 1.9\% & 11.1\% & 8.2\% & 7.1\% & 4.0\% & 10.1\% & 3.9\% & 8.3\% & 7.3\% & 8.8\% & 9.1\% & 4.1\% & 4.0\% & \databar{5.4}{100}{failred}{5.4\%} \\
    C4.1. over\_broad\_change & redaction too broad, breaks legitimate callers & \tagchip{scopegreen}{\infoex} & 2.5\% & 9.3\% & 4.6\% & 4.3\% & 3.3\% & 3.4\% & 2.5\% & 1.8\% & 0.8\% & 0.9\% & 9.7\% & 8.2\% & 6.0\% & 2.7\% & 10.1\% & 3.9\% & 6.9\% & 6.1\% & 6.2\% & 7.8\% & 2.7\% & 2.7\% & \databar{4.5}{100}{failred}{4.5\%} \\
    \rowcolor{black!4}C4.2. wrong\_value\_regression & changed value or fallback breaks other behaviour & \tagchip{scopegreen}{\infoex} & 0.1\% & 0.2\% & 0.1\% & 0.2\% & 0.0\% & 0.0\% & 0.0\% & 0.0\% & 0.0\% & 0.0\% & 0.0\% & 0.0\% & 1.2\% & 0.0\% & 0.0\% & 0.0\% & 0.0\% & 0.0\% & 1.2\% & 0.0\% & 0.0\% & 0.0\% & \databar{0.1}{100}{failred}{0.1\%} \\
    C4.3. other\_regression & other behaviour stops working after the fix & \tagchip{scopeblue}{any\ fix} & 1.0\% & 0.0\% & 1.1\% & 0.0\% & 0.0\% & 0.0\% & 1.2\% & 0.9\% & 0.0\% & 0.9\% & 1.4\% & 0.0\% & 0.0\% & 1.3\% & 0.0\% & 0.0\% & 1.4\% & 1.2\% & 1.2\% & 1.3\% & 1.4\% & 1.3\% & \databar{0.7}{100}{failred}{0.7\%} \\
    \rowcolor{black!4}\textbf{C5. non\_termination} & the patched code hangs, evaluation cannot finish & \tagchip{scopeblue}{any\ fix} & 0.4\% & 0.0\% & 0.4\% & 0.0\% & 0.0\% & 0.0\% & 0.0\% & 0.9\% & 0.0\% & 0.9\% & 1.4\% & 0.0\% & 0.0\% & 0.0\% & 0.0\% & 0.0\% & 0.0\% & 0.0\% & 0.0\% & 0.0\% & 1.4\% & 0.0\% & \databar{0.3}{100}{failred}{0.3\%} \\
    \midrule
    \textit{Analysed failures} & & & 100.0\% & 100.0\% & 100.0\% & 100.0\% & 100.0\% & 100.0\% & 100.0\% & 100.0\% & 100.0\% & 100.0\% & 100.0\% & 100.0\% & 100.0\% & 100.0\% & 100.0\% & 100.0\% & 100.0\% & 100.0\% & 100.0\% & 100.0\% & 100.0\% & 100.0\% & 100.0\% \\
    \bottomrule
  \end{tabular}}
\end{table}

%% file: 7_ThreatsToValidity.tex
\section{Threats to Validity}
\label{sec:threats}

\textbf{Training-data contamination.}
Because LLMs may have seen the public GitHub data underlying {\bench} during training, we split the benchmark at February 16, 2026, the latest knowledge cutoff among the three evaluated LLMs, into 105 pre-boundary and 56 post-boundary instances and report both separately. Contamination may still affect the pre-boundary subset.

\noindent \textbf{Unintended external information access.}
Agentic systems may access unavailable information through Web search or future Git history. We mitigate this threat by blocking all network access except LLM API requests and removing future commits. However, LLM API providers may offer server-side Web search that bypasses this isolation. It was initially enabled for Codex with Qwen3.5-Plus, yielding 64 resolved defects versus 58 after disabling it. We corrected the setting for subsequent runs and inspected all trajectories, finding no other such access.

\noindent \textbf{Test-oracle validity.}
Developer-written \texttt{FAIL\_TO\_PASS} tests may encode reference-fix details rather than intended security outcomes. We therefore inspect all tests, fix implementation-specific assertions in 57 of 161 instances, and verify each final exposure test on both buggy and fixed versions. Nevertheless, finite tests may miss exposure paths or regressions, motivating future \infoex-specific methods for assessing whether all relevant exposure paths are blocked.

\noindent \textbf{Benchmark representativeness.}
{\bench} focuses on Top-ranked Python and Java repositories and defects with merged source-code fixes and executable regression tests, and may not represent other languages, projects, or repair settings. Future work can extend it to broader scenarios.

\noindent \textbf{Benchmark construction and annotation.}
LLM pre-screening may miss relevant defects, while manual review and characterization involve subjective judgment. We mitigate the former by assessing each candidate three times, retaining it if any assessment identifies an exposure, and auditing 600 pairs rejected by all three, finding one missed \infoex defect. For manual review, two authors independently assess retained candidates using evidence of sensitive information, exposure channel, and unauthorized actor, resolving disagreements through discussion. The six characterization dimensions are likewise developed and annotated by two authors.

%% file: 8_Conclusion.tex
\section{Conclusion}
\label{sec:conclusion}

We present {\bench}, the first benchmark for automated repair of information exposure defects, with 161 execution-verified instances from 64 Python and Java repositories. Across six repair systems and three LLMs, the best configuration resolves 39.1\% of the instances, while 40.4\% remain unresolved by all 18 configurations. Most failures involve incomplete exposure removal, highlighting the need to close every path through which the sensitive information can still be disclosed. We hope {\bench} provides a foundation for systematically evaluating and advancing future \infoex repair techniques.